\documentclass[
 reprint,
 superscriptaddress,
 amsmath,amssymb,
 aps,
pra,
showkeys
]{revtex4-2}

\newcommand{\SmallUpperCase}[1]{\textsc{\MakeLowercase{#1}}}
\usepackage{soul}
\usepackage{float}
\usepackage{multibib}
\usepackage{hyperref}
\usepackage{xcolor}
\hypersetup{
    colorlinks,
    linkcolor={blue!50!blue},
    citecolor={blue!50!blue},
    urlcolor={blue!80!black}
}
\usepackage{siunitx}
\usepackage{import}
\usepackage{placeins}
\usepackage{xcolor}
\usepackage{lmodern}
\usepackage[braket, qm]{qcircuit}
\usepackage{graphicx}
\usepackage{dcolumn}
\usepackage{bm}
\usepackage{comment}
\usepackage[english]{babel}
\usepackage{blindtext}
\usepackage{physics}  
\usepackage{multirow}
\usepackage{array}

\begin{document}

\title{Comprehensive scheme for identifying defects in solid-state quantum systems}

\author{Chanaprom Cholsuk}
\email{chanaprom.cholsuk@uni-jena.de}
\affiliation{Abbe Center of Photonics, Institute of Applied Physics, Friedrich Schiller University Jena, 07745 Jena, Germany}

\author{Sujin Suwanna}
\affiliation{Optical and Quantum Physics Laboratory, Department of Physics, Faculty of Science, Mahidol University, Bangkok 10400, Thailand}.

\author{Tobias Vogl}%
\email{tobias.vogl@uni-jena.de}
\affiliation{Abbe Center of Photonics, Institute of Applied Physics, Friedrich Schiller University Jena, 07745 Jena, Germany}
\affiliation{Fraunhofer-Institute for Applied Optics and Precision Engineering IOF, 07745 Jena, Germany}

\date{\today}

\begin{abstract}
A solid-state quantum emitter is a crucial component for optical quantum technologies, ideally with a compatible wavelength for efficient coupling to other components in a quantum network. It is essential to understand fluorescent defects that lead to specific emitters. In this work, we employ density functional theory (DFT) to demonstrate the calculation of the complete optical fingerprints of quantum emitters in hexagonal boron nitride. Our results suggest that instead of comparing a single optical property, like the zero-phonon line energy, multiple properties should be used when comparing simulations to the experiment. Moreover, we apply this approach to predict the suitability of using the emitters in specific quantum applications. We therefore apply DFT calculations to identify quantum emitters with a lower risk of misassignments and a way to design optical quantum systems. Hence, we provide a recipe for classification and generation of universal quantum emitters in future hybrid quantum networks.
\end{abstract}

\keywords{single photon emitter, hexagonal boron nitride, density functional theory, defect identification, quantum technology applications}

\maketitle

\section{Introduction}
Many solid-state single photon emitters (SPEs) originate from local modifications to the electronic structure of a crystal material. Relatively simple examples are point-like defects that introduce additional energy levels into the band gap of the crystal or strain-induced band bending. Examples include color centers in diamond \cite{Bradac2019} and silicon carbide \cite{doi:10.1126/sciadv.abj0627}. Due to supporting an intrinsically high photon extraction efficiency \cite{Vogl2019}, two-dimensional (2D) materials such as transition metal dichalcogenides (TMDs) \cite{spe-tmd,srivastava_sidler_allain_lembke_kis_imamoglu_2015} and hexagonal boron nitride (hBN) \cite{Tran2016, vogl2018} have become a promising system to realize a single photon source. The ability to identify such fluorescent defects is crucial for quantum technology applications as it allows one to reveal and manipulate their photophysical properties.\\
\indent For instance, optical quantum computing with quantum emitters requires the two-level states of a defect to act as a qubit that logic gates can be performed on \cite{doi:10.1126/sciadv.abj0627, Brien, Reilly2015, Pezzagna2021, Conlon2023}. For quantum key distribution, some wavelength yields more efficient links, i.e., in the telecom band for fiber \cite{Chapuran2009} or specific visible wavelengths for free-space connections \cite{mostafa-qkd}. On the other hand, quantum sensing needs different defect characteristics, such as singlet and triplet spin configurations to enable optically detected magnetic resonance (ODMR) for magnetometry \cite{Aslam2023}. In the future, we will not only have single quantum computers and point-to-point quantum communication but a large quantum network with distributed remote sensing. These will involve memories that the emitters should couple to efficiently \cite{quick3}. In addition, when using emitters as memories, they need a specific level structure (e.g., a three-level $\Lambda$ structure consisting of ground-, excited-, and meta-stable states) \cite{Lvovsky2009,Pingault2014}.\ This suggests that when designing defect emitters, all photophysical properties need to be considered to identify whether the defect candidate is suitable for the desired application.\\
\indent Focusing on room-temperature operation, fluorescent defects in hBN are considered to outperform others in the sense of robustness against aging \cite{vogl2018,LIU2020114251}, radiation \cite{Vogl2019-radiation}, temperature \cite{Kianinia2017}, while at the same time having a high single photon luminosity \cite{Tran2016}, featuring pure single photon emission \cite{anand-dipole}, and a short excited state lifetime \cite{vogl2018}. It is well known that defects in hBN can emit over a broad range of wavelengths \cite{Cholsuk2022}, including the UV \cite{Cdimer-4eV}, in the visible spectrum \cite{Dietrich2018}, as well as in the near-infrared \cite{Camphausen2020}. Consequently, assigning the microscopic origin of hBN defect for a specific emission wavelength becomes a challenging task. Several strategies have been carried out to compare simulations with the experiments \cite{Grosso2020,gali2018,Sajid2020-VNNB,Auburger2021,Wu2019}. A common method for SPE identification is by comparing zero-phonon line (ZPL) energy and photoluminescence (PL) lineshape with first-principle calculations, generally based on density functional theory (DFT). Recently, one possibility of a 2 eV emitter has been attributed to the C$_\text{B}$V$_\text{N}$ defect by matching the electronic transition and its experimental PL spectrum \cite{Sajid2020}. Similarly, carbon-based defects up to tetramers were investigated by comparing the phonon side-band (PSB) instead of their ZPL \cite{Jara2020}. They found that C$_\text{B}$C$_\text{N}$C$_\text{N}$ has quite similar PL characteristics, but needs some considerable redshift of the ZPL to match the experimental data. Another DFT study further considered Huang-Rhys factor (HR) and lifetime of the second electronic transition of the C$_\text{B}$C$_\text{N}$C$_\text{N}$ defect \cite{Li2022}. These identification techniques have also been employed for calculating the wavelengths of other hBN emitters \cite{Cdimer-4eV}, as well as for characterizing their electronic structures \cite{Smart2021,Sajid2018,VB-1-Gali}.\\
\indent Among several DFT studies, it has been realized that some parameters agree well with the experiment whereas other parameters are inconsistent. This could stem from several reasons: (i) strain in the local crystal environment can significantly alter the photophysical properties and this effect is amplified in a 2D crystal (compared to 3D); (ii) complicated coupling terms of the defect, which simulation sometimes cannot capture well; or (iii) misassignment of the defect with another one that has similar optical properties. Recent works tackled this issue by proposing another method to determine the defect from electron paramagnetic resonance (EPR) and found agreement between DFT and experiment \cite{Li2022-oxygen, Sajid2018-EPR}. It becomes clear that defect identification should not rely on only one property. In other words, the comprehensive optical fingerprint needs to be considered when determining the defect's properties and comparing to the experiment. This way, the three issues mentioned above can be avoided with a much higher probability.\\
\indent In this work, we demonstrate calculating the complete fingerprint of hBN quantum emitters. In particular, the proposed identification method only requires the common experimentally measured properties for the comparison which can be obtained using a standard photoluminescence microscope. This methodology is showcased with the example of the C$_\text{B}$C$_\text{N}$C$_\text{B}$C$_\text{N}$ defect, which is consistent to our previous experiments \cite{anand-yellow, anand-dipole}, as well as with the C$_\text{B}$C$_\text{N}$C$_\text{N}$, which was proposed earlier \cite{Jara2020, Li2022}. In addition, we also apply the method to predict the full fingerprint of the Al$_\text{N}$ and P$_\text{N}$V$_\text{B}$ defects. These have been selected as promising candidates for quantum technology applications based on their ZPL energy before \cite{Cholsuk2022}. Which actual applications are possible is studied in this work.
\section{Methodology}
\subsection{Ground and excited structural relaxation}
\indent All structures have been simulated using spin-polarized DFT calculations with the Vienna Ab initio Simulation Package (VASP) employing a plane wave basis set \cite{vasp1,vasp2} and the projector augmented wave (PAW) method for the pseudopotentials \cite{paw,paw2}. All geometry relaxations were performed with an energy cutoff at 500 eV and the total energy convergence with an accuracy of 10$^{-4}$ eV. To prevent the interaction from neighboring cells, we added a 15 \AA$ $ vacuum layer and used a supercell size of $7\times 7\times 1$ containing 98 atoms (pristine). To avoid underestimating the band gap, the HSE06 functional was employed for all calculations with the single $\Gamma$-point scheme to enable relaxation of the structures with only internal coordinates allowed until all forces were lower than 0.01 eV/\AA. This yields the calculated band gap of around 6 eV as shown in the results, similar to other reports \cite{Abdi2018,Sajid2018-EPR}. For the excited-state calculations, the $\Delta$SCF method was employed to manually occupy electrons for the excited-state configuration \cite{Jones1989}. Note that the total energy of both ground and excited state configurations was compared among singlet, doublet, and triplet states. Only the most stable configuration was selected for further analysis. It should be remarked that every DFT strategy has only finite accuracy, so comparing the prediction calculated by different strategies is essential, as recently pointed out \cite{Reimers2020-VB-1}. For the HSE06 functional with the periodic model, when comparing with other DFT techniques, it still qualitatively agrees well unless a defect includes multireference character \cite{doi:10.1021/acs.jctc.7b01072,Sajid2020-VNNB}. In particular, HSE06 has demonstrated very good agreement with experiments in comparison with the generalized gradient approximation \cite{doi:10.1021/acs.jctc.7b01072}.
\begin{figure*}[ht]
    \centering
    \includegraphics[width = 0.85\textwidth]{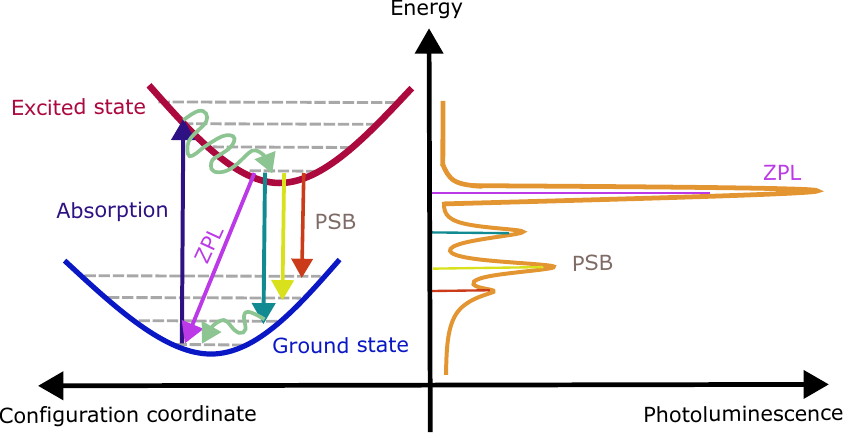}
    \caption{A schematic illustration of recombination mechanism between occupied (ground) and unoccupied (excited) defect states and the resulting emission spectrum that one observes in an experiment. After (off-resonant) excitation, the radiative transition is via ZPL or PSB paths, both of which can be observed in PL lineshape, whereas non-radiative transitions (phonon scattering) are via the wavy light green arrows.
    }
    \label{fig:recombination}
\end{figure*}
\subsection{Transition dipole moment}
\indent For the defect-defect transition, we focus on two dipoles existing during the recombination process as depicted in Fig.\ \ref{fig:recombination}, namely the excitation and emission dipoles. While both dipoles can be identified by group symmetry of each defect, most defects, especially some defect complexes, encounter broken structural symmetry, impeding the defect assignment by group theory. As a result, we need to investigate the dipoles explicitly for each individual defect. The transition dipole moment (TDM) $\mu$ can be calculated using the wavefunctions of the initial/final states $\psi_{i/f}$ with
\begin{equation}
\boldsymbol{\mu} = \frac{i\hbar}{(E_{f} - E_{i})m}\bra{\psi_{f}}\textbf{p}\ket{\psi_{i}},
\label{eq:dipole}
\end{equation}
where $E_{i/f}$ are the eigenvalues of the initial/final orbitals, respectively, $m$ is the electron mass; and $\mathbf{p}$ is a momentum operator.
For the excitation TDM, we define $\psi_{i}$ as the most stable ground-state configuration (the lowest total energy of the blue curve in Fig.\ \ref{fig:recombination}), whilst $\psi_{f}$ as the most stable excited-state configuration (the lowest total energy of the red curve in Fig.\ \ref{fig:recombination}). For the emission TDM, the roles of $\psi_{i}$ and $\psi_{f}$ are reversed. Note that $\psi_{f}$ from the excitation TDM and $\psi_{i}$ from the emission TDM in general are not identical. The PyVaspwfc Python code was used to extract the wave functions \cite{pyWave}, whereas its modified version was used to handle wave functions from the ground and excited states \cite{Davidsson2020}. Equivalently, the transition dipole can be expressed as
\begin{equation}
\boldsymbol{\mu} = \abs{\mu_x}\hat{x} + \abs{\mu_y}\hat{y} + \abs{\mu_z}\hat{z},
\end{equation}
where $\mu_z$ = 0 indicates a purely \textit{in-plane} dipole; otherwise, it has an \textit{out-of-plane} contribution. Note that in our coordinate system, the $xy$-plane is the hBN crystal plane. As the excitation/emission polarizations are orthogonal to their dipoles, we rotate both dipoles by 90$^\circ$ and take modulo 60$^\circ$ to obtain the angle with respect to the nearest crystal axis. Of course, the hexagonal crystal lattice is spaced 120$^\circ$; however, after 180$^\circ$ the angle from the crystal axis will be identical. Finally, both dipoles were also projected onto the $xy$-plane to calculate the in-plane polarization visibility. This makes our theoretical results directly compatible to compare with experiments of polarization-resolved PL.
\subsection{Recombination process}
\indent Characteristics of the excitation/emission process are also the temporal dynamics. The radiative transition rate is an estimator of the desired decay pathway. This quantity can be computed by
\begin{equation}
\Gamma_{\mathrm{R}}=\frac{n_D e^2}{3 \pi \epsilon_0 \hbar^4 c^3} E_0^3 \mu_{\mathrm{e}-\mathrm{h}}^2,
\end{equation}
where $e$ is the electron charge; $\epsilon_0$ is vacuum permittivity; $E_0$ is the transition energy defined as the energy difference between ground and excited states; $n_{\mathrm{D}}$ is the refractive index of the host crystal (for hBN this is 1.85 in the visible \cite{vogl2018}); and $\mu_{\mathrm{e}-\mathrm{h}}^2$ is the modulus square of dipole moment obtained by Eq.\ \ref{eq:dipole}. It is worth noting that in most experiments, the hBN layers are attached to a substrate (often SiO$_2$), which can reduce or enhance the density of states that the emitter can couple to. This can therefore modify the dipole emission pattern and affect the emitter lifetime (and therefore also radiative transition rate) due to the Purcell effect \cite{Vogl2019}. This cannot be calculated using DFT; however, the lifetime calculated with DFT can be scaled with the Purcell factor calculated with other techniques (e.g., Finite-Difference Time-Domain (FDTD) simulations) to take into account the dielectric environment. We demonstrate this in the Supplementary Section S1.\\
\indent Apart from the radiative decay pathway, the excitation can also relax non-radiatively to the ground state through phonon-assisted recombination. We compute the rate of this process using the \SmallUpperCase{Nonrad} Python code \cite{Turiansky2021,alkauskas_first-principles_2014} developed based on Fermi's golden rule formalism, that is
\begin{equation}
\Gamma_{\mathrm{NR}}=\frac{2 \pi}{\hbar} g \sum_{n, m} p_{i n}\left|\left\langle f m\left|H^{\mathrm{e}-\mathrm{ph}}\right| i n\right\rangle\right|^2 \delta\left(E_{f m}-E_{i n}\right),
\label{eq:fermi-golden}
\end{equation}
where $\Gamma_{\mathrm{NR}}$ is the non-radiative transition rate between electron state $i$ in phonon state $n$ and electron state $f$ in phonon state $m$. $g$ is the degeneracy factor. $p_{\text{in}}$ is the thermal probability distribution of state $|i n\rangle$ following the Boltzmann distribution. $H^{\mathrm{e}-\mathrm{ph}}$ is the electron-phonon coupling Hamiltonian. In practice, Eq.\ \ref{eq:fermi-golden} can be simplified by the static coupling approximation with the one-dimensional phonon approximation, so we obtain
\begin{eqnarray}
 &\Gamma_{\mathrm{NR}}=\frac{2 \pi}{\hbar} g\left|W_{i f}\right|^2 X_{i f}(T), \\
& W_{i f}  =\left.\left\langle\psi_i(\mathbf{r}, \mathbf{R})\left|\frac{\partial H}{\partial Q}\right| \psi_f(\mathbf{r}, \mathbf{R})\right\rangle\right|_{\mathbf{R}=\mathbf{R}_{\mathbf{a}}}, \\
& X_{i f}=  \sum_{n, m} p_{i n}\left|\left\langle\phi_{f m}(\mathbf{R})\left|Q-Q_a\right| \phi_{i n}(\mathbf{R})\right\rangle\right|^2 \times \nonumber  \\
&\delta\left(m \hbar \omega_f-n \hbar \omega_i+\Delta E_{i f}\right),
\end{eqnarray}
where the phonon term $X_{i f}$ and the electronic term $W_{i f}$ become separated. More detail can be found in the work of Turiansky \emph{et al.}\cite{Turiansky2021}.\\
\indent Considering both radiative and non-radiative transitions, we are able to estimate the quantum efficiency $\eta$ of an emitter by the following equation
\begin{equation}
    \eta = \frac{\Gamma_{\mathrm{R}}}{\Gamma_{\mathrm{R}}+\Gamma_{\mathrm{NR}}}. \label{eq:q_eff}
\end{equation}
\begin{table*}[htb]
\begin{tabular}{c |c c c| c c | c}
\hline
\multirow{2}{*}{Photophysical}  & \multicolumn{6}{c}{ Defects } \\
 & \multicolumn{3}{c}{C$_{2}$C$_\text{N}$}  & \multicolumn{2}{c}{C$_{2}$C$_\text{N}$} &  C$_{2}$C$_{2}$ \\
Properties   & Ours & Ref.\cite{Jara2020} & Ref.\cite{Li2022} & Ours & Ref.\cite{Li2022} & Ours \\
 & HSE06 & HSE06 & GW & HSE06 & GW & HSE06 \\
\hline
Transition order & 1st & 1st & 1st & 2nd & 2nd & 1st
\\Most stable config. & doublet & doublet & doublet & doublet & doublet & singlet \\
 Spin transition & $\downarrow$ - $\downarrow$ & $\downarrow$ - $\downarrow$ & $\downarrow$ - $\downarrow$ & $\downarrow$ - $\downarrow$ & $\downarrow$ - $\downarrow$ & $\downarrow$ - $\downarrow$ \\
ZPL (nm) &  751   & 765 & 1025 & 495 & 582  &573  \\
 E$_0$ (eV) & 1.65  &  1.62 & 1.21  &   2.50   &  2.13 &2.16 \\
 $\Delta Q$ (amu$^{1/2}$\AA) &  0.43   &  - & 0.19 & 0.39 & 0.27 & 0.44 \\
 HR &  0.97 & -  & 0.95 & 0.85 & 1.35 & 1.00 \\
 DW &  0.38 & -  & 0.39 &  0.43 & 0.26 &  0.37   \\
  Excitation & \multirow{2}{*}{0.05} & \multirow{2}{*}{-}  & \multirow{2}{*}{-} & \multirow{2}{*}{0.49} & \multirow{2}{*}{-} & \multirow{2}{*}{11.12} \\
 polarization ($^\circ$) &     &   &  &  &  \\
 Excitation  &   \multirow{2}{*}{1.00}  & \multirow{2}{*}{-}  & \multirow{2}{*}{-} & \multirow{2}{*}{1.00} & \multirow{2}{*}{-} & \multirow{2}{*}{1.00} \\
 In-plane visibility & & & & & \\
Emission  & \multirow{2}{*}{1.54}  &   \multirow{2}{*}{-}  & \multirow{2}{*}{-}  & \multirow{2}{*}{18.77} & \multirow{2}{*}{-} & \multirow{2}{*}{12.14}   \\
polarization ($^\circ$) & & & & & \\
 Emission  &  \multirow{2}{*}{1.00}  &   \multirow{2}{*}{-}  & \multirow{2}{*}{-}  & \multirow{2}{*}{1.00} & \multirow{2}{*}{-} & \multirow{2}{*}{1.00}  \\
 In-plane visibility & & & & & \\
 $\Gamma_R$ (1/s) & $6.38\times 10^7$  &  -  & $1.26\times 10^6$  & $2.42\times 10^7$ &  $1.93\times 10^7$ & $5.90\times 10^7$ \\
 $\tau_R$ (ns)& 15.67 &  -  & 795.00   & 41.36  & 51.90  & 16.94\\
 $\Gamma_{NR}$ (1/s) & $6.99\times10^2$ & - & $\sim1\times10^3$ & $7.47\times10^5$ &  $4.37\times10^8$ & $8.53\times10^8$\\
 $\tau_{NR}$ (ns) & $1.43\times10^6$ & - & $\sim1\times10^6$ & $1.34\times10^3$ & 2.29 & 1.17 \\
 $\eta$ (\%) & $\sim100.00$ & - & - & 97.00 & 4.23$^{*}$ & 6.47 \\
ODMR & unlikely & - & - & unlikely & - & unlikely  \\
\hline
\multicolumn{4}{l}{\footnotesize{* This quantum efficiency was calculated differently from Eq.~\ref{eq:q_eff}.}}\\
\end{tabular}
\caption{Complete optical fingerprints of the C$_{2}$C$_\text{N}$ and C$_{2}$C$_{2}$ defects as candidates for 2 eV emitter in hBN. The arrow $\downarrow$ - $\downarrow$ indicates a transition preserving spin-down.}
  \label{tab:list_parameters}
\end{table*}
\subsection{Photoluminescence}
\indent Additional optical properties were calculated using the PyPhotonics workflow \cite{Tawfik2022} with the theory based on the work of Alkauskas \emph{et al.} \cite{Alkauskas2014}. For the sake of completeness, we summarize the workflow in the following. After constraining the electron occupation using the $\Delta$SCF formalism \cite{Jones1989}, we can evaluate how the excited-state configuration is displaced from the ground-state one using the configuration coordinate $q_{k}$:
\begin{equation}
    q_k = \sum_{\alpha,i}\sqrt{m_\alpha}\left(R_{e,\alpha i} - R_{g,\alpha i}\right)\Delta r_{k,\alpha i},
\end{equation}
where $\alpha$ and $i$ run over the atomic species and the spatial coordinates $(x,y,z)$, respectively. $m_{\alpha}$ is the mass of atom species $\alpha$. $R_g$ and $R_e$ are the stable atomic positions in the ground and excited states, accordingly, while $\Delta r_{k,\alpha i}$ is the displacement vector of atom $\alpha$ at the phonon mode $k$ between the ground and excited states.\\
\indent Then, to explore the phonon coupling, the partial Huang-Rhys (HR) factor $s_k$ for each phonon mode $k$ is computed by
\begin{equation}
    s_k = \frac{\omega_k q_k^2}{2\hbar}.
\end{equation}
This quantity leads to the total HR factor $S$ as shown in the following equation.
\begin{equation}
    S(\hbar\omega) = \sum_k s_k\delta(\hbar\omega - \hbar\omega_k).
\end{equation}
Note that the Debye-Waller (DW) factor can also be calculated here by $\exp(-S)$. To obtain the PL spectrum, we first compute the time-dependent spectral function $S(t)$ given by
\begin{equation}
    S(t) = \int_0^\infty S(\hbar\omega)\exp(-i\omega t)d(\hbar\omega).
\end{equation}
Finally, the PL intensity can be obtained from
\begin{equation}
    L(\hbar\omega) = C\omega^3 A(\hbar\omega),
\end{equation}
where $C$ is a normalization constant (usually obtained by fitting experimental data), and $A(\hbar\omega)$ is the optical spectral function computed by
\begin{equation}
    A(E_{ZPL} - \hbar\omega) = \frac{1}{2\pi}\int_{-\infty}^{\infty}G(t)\exp(-i\omega t-\gamma|t|) dt,
\end{equation}
where $G(t)$ is the generating function of $G(t) = \exp(S(t) - S(0))$, and $\gamma$ is a fitting parameter. Note that the convergence of the PL spectrum for all investigated defects was checked by varying the supercell size as shown in the Supplementary Section S2.\\
\section{Results and discussion}
\subsection{Parameters for defect identification}
\indent We now showcase the methodology outlined above and attempt to identify the 2 eV emitter by considering the entire optical fingerprint. It is already known that the ZPL calculated with DFT depends on the specific functional and basis set used \cite{doi:10.1021/acs.jctc.7b01072}. This actually applies to all electronic properties, requiring a careful choice of the computational setup. The most accurate calculations can be performed using hybrid HSE06 or GW approximations. Many experimental and theoretical studies have speculated about carbon playing a role for the 2 eV emitter \cite{Mendelson2021,anand-dipole,anand-yellow,Wang2018,Exarhos2017,Kozawa2023}. In the following, C$_\text{B}$C$_\text{N}$ is abbreviated as C$_2$. Among the most promising candidates, the carbon trimer \cite{Jara2020,Li2022} C$_{2}$C$_\text{N}$ and tetramer \cite{anand-yellow,anand-dipole} C$_{2}$C$_{2}$ in a specific configuration have been proposed. Note that not all works consider the first-order transition between highest occupied and lowest unoccupied state but also second- and higher orders \cite{Li2022,Jungwirth2017}. These cases cannot be distinguished easily by comparing the spectra and it is required to compare the decay pathways.\\
\indent The complete fingerprints of these candidates are shown in Tab.\ \ref{tab:list_parameters}. Note that optically detected magnetic resonance (ODMR) listed indicates the possibility of having this feature by considering which spin configuration without an external magnetic field is the most stable. In principle, for exhibiting ODMR, a defect must inherit a decay pathway via the intersystem crossing channel where the triplet spin configuration acts as the preferred configuration, while the singlet spin configuration acts as the meta-stable state. With the triplet spin configuration, the sub-level states can be split into the $m_{s}=0,\pm1$ states due to zero field splitting even without an external magnetic field. This is one of the crucial requirements for the ODMR effect. As such, we exploited this fact and hypothesized that the triplet state needs to be most stable in order to have this state splitting. We applied this criterion to determine the well-studied V$_\text{B}^{-1}$ and predict with our method that this defect is likely ODMR-active, consistent with previous studies \cite{VB-1-Gali}. However, it is worth noting that further investigation of the defect state splitting by applying an external field is still needed and is part of future work.\\
\begin{figure*}[ht!]
    \centering
    \includegraphics[width = 1\textwidth]{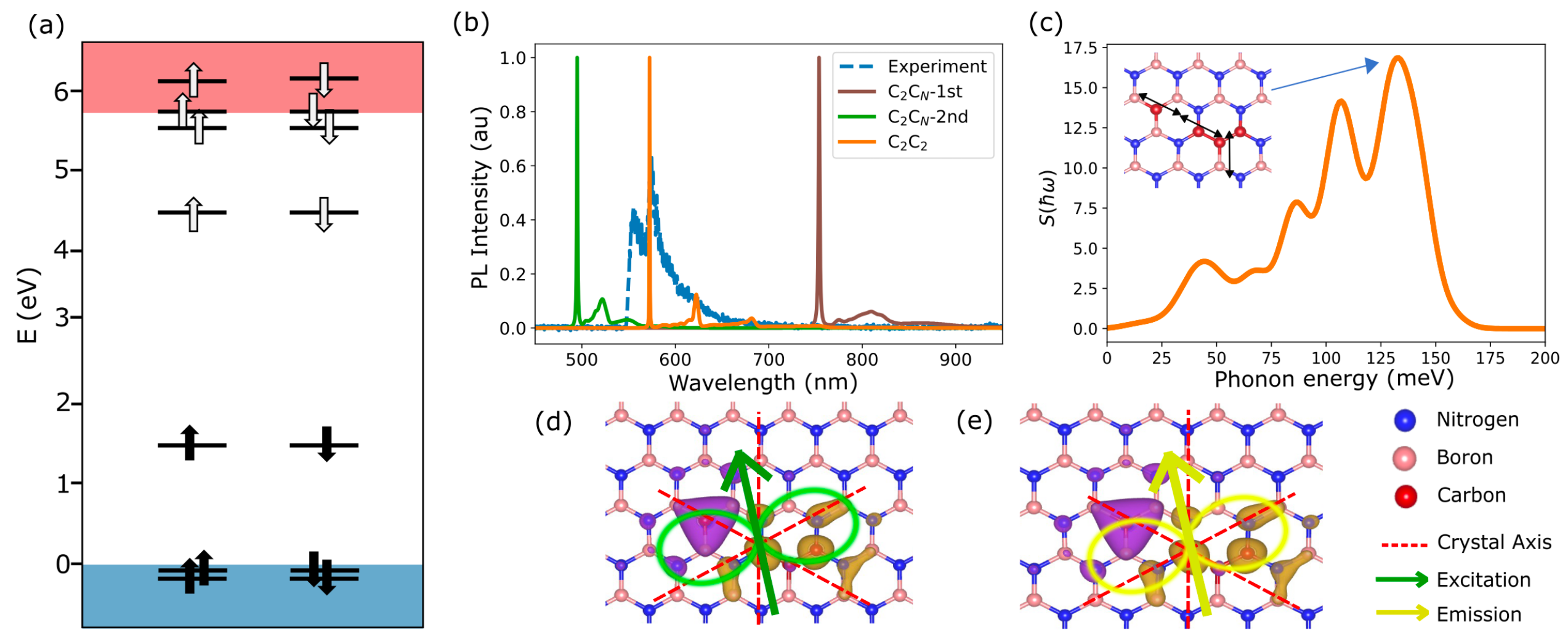}
    \caption{Photophysical properties of the C$_2$C$_2$ defect. (a) Single-particle electronic transition where the filled/unfilled arrows identify occupied/unoccupied states. The brown and blue shaded areas illustrate the valence and conduction bands, respectively. The band gap is extracted to be 5.84 eV. (b) Photoluminescence spectrum where the blue, brown, green, and orange lines depict the experimental data, as well as calculated transitions of C$_2$C$_\text{N}$ (1$^{st}$ and 2$^{nd}$ transitions), and C$_2$C$_2$, respectively. (c) Spectral function of C$_2$C$_2$ defect where the structure is relaxed by HSE06 functional. (d) The excitation dipole orientation of C$_2$C$_2$. (e) The emission orientation of C$_2$C$_2$. The arrow indicates the dipole direction whereas the circular shape displays the polarization. The isosurface represents the charge density difference between ground and excited states. Purple and yellow isosurfaces depict negative and positive charges. Data of C$_2$C$_2$ is adapted from an experiment \cite{anand-dipole}.
    }
    \label{fig:c2c2}
\end{figure*}
\subsection{Identifying defects for 2 eV emitters}
\indent Fig.\ \ref{fig:c2c2}(a) depicts the most stable configuration for C$_{2}$C$_{2}$ as a singlet, which allows either spin up or down for the transition. Assuming one spin-down electron is excited, after electron relaxation, it will relax via the ZPL at 573 nm or PSB as shown in Fig.\ \ref{fig:c2c2}(b). Note that when comparing the DFT-calculated spectrum to that taken from an experiment \cite{anand-yellow,anand-dipole}, no temperature broadening mechanism was considered. For comparison, we also show the spectrum of the C$_{2}$C$_\text{N}$ trimer. The phonon contribution is relatively weak with an HR factor around 1 consistent with the experimental range between 0.70 and 1.70 \cite{Wang2018,Exarhos2017}, and originates from the phonon mode as illustrated in the inset of Fig.\ \ref{fig:c2c2}(c). For the decay channel, the radiative lifetime is estimated to be 16.94 ns whereas the non-radiative lifetime is roughly 1.17 ns, leading to the quantum efficiency of around 6.47\%. The experimental lifetime was measured to be around 4 ns \cite{anand-yellow}, while the quantum efficiency (or non-radiative decay) has not been estimated. Based on the emitter brightness compared to the brightness of other emitters with a measured quantum efficiency in a similar setup, we estimate the experimental quantum efficiency to be on the order of a few percent, consistent with the simulations. Then, for the dipole orientation, our calculated excitation and emission polarizations make 11.12$^\circ$ and 12.14$^\circ$ away from the crystal axis, respectively, as shown in Fig.\ \ref{fig:c2c2}(d) and (e). The excitation polarization fits well with the experimental angle ranging 3.90$^\circ$ to 11.57$^\circ$, while the emission one is slightly off with the experimental angles of 16.24$^\circ$ and 20.73$^\circ$ away from the nearest crystal axis \cite{anand-dipole}. There is still a large uncertainty in the experiment that needs to be improved for a better comparison. The degree of polarization, however, matches well with the experimentally observed well-polarized light.\\
\indent We can directly compare these results of the C$_{2}$C$_{2}$ defect with the C$_{2}$C$_{\text{N}}$. Our calculated ZPL is 1.65 eV and 2.50 eV for the first and second electronic transitions, which are in excellent agreement with previous HSE06 studies reporting at 1.64 eV and 2.41 eV \cite{Jara2020,Li2022}. For the lifetimes, we estimated 15.7 ns and 41.4 ns for the first and second transitions, respectively. The excitation (emission) polarization of this defect is found to be in-plane and yields 0.05$^\circ$ (1.54$^\circ$) and 0.49$^\circ$ (18.77$^\circ$) for the first and second electronic transitions, respectively. This is actually quite consistent with the experimental data \cite{anand-dipole}; however, the deviation in other photophysical properties makes it unlikely to be responsible for this particular 2 eV emitter \cite{anand-dipole,anand-yellow}. Considering the ability to exhibit ODMR, both defects are unlikely to inherit this feature as the most stable ground-state configuration is not triplet; however, investigation after including the magnetic field is required. We emphasize that we are not arguing that  C$_{2}$C$_{\text{N}}$ is not likely for 2 eV emitter, but rather suggesting that C$_{2}$C$_{2}$ can be one another possible candidate. As such, realizing dipole orientation together with other optical properties can enhance the certainty of defect identification, as can be seen from C$_2$C$_2$ and C$_{2}$C$_{\text{N}}$ defects.\\
\subsection{Defects suitable for NV center coupling and daylight quantum communication}
\indent By applying the same procedure as for defect identification, we can also calculate the full photophysical fingerprint of novel defects and judge their suitability for quantum technologies. We have previously done this with ZLP energy alone \cite{Cholsuk2022}. However, even if the wavelength is suitable, an out-of-plane oriented dipole or a very long excited-state lifetime will prevent usage in applications such as quantum communication. The defects we study are Al$_\text{N}$ and P$_\text{N}$V$_\text{B}$. The former is predicted to be useful for coupling with NV centers in diamond \cite{Cholsuk2022}, while the latter is predicted to be suitable for free-space daylight quantum communication \cite{mostafa-qkd}. In such scheme, a quantum light source is needed at one of the Fraunhofer lines in the solar spectrum, and if one filters narrowly around this line in a free-space receiver, there is no solar background and quantum information can be transmitted in ambient conditions in roof-to-roof or space-to-ground scenarios.\\
\begin{table}[hbt]
\begin{tabular}{c c c}
\hline
Photophysical & \multicolumn{2}{c}{ Defects } \\
 Properties &  Al$_\text{N}$ & P$_\text{N}$V$_\text{B}$ \\
\hline 
Transition order  & 1st & 1st 
\\Most stable configuration  & Singlet & Doublet \\
 Spin transition  & $\uparrow$ - $\uparrow$ & $\downarrow$ - $\downarrow$ \\
ZPL (nm) & 682 & 673 \\
 $\Delta Q$ (amu$^{1/2}$\AA) &0.83 & 0.70 \\
 HR &     2.94 & 1.87 \\
 DW &     0.05 & 0.15 \\
 Excitation polarization ($^\circ$) &   9.66 & 5.02 \\
 Excitation In-plane visibility &   0.61 & 1.00 \\
Emission polarization ($^\circ$) &    5.43 & 16.85 \\
 Emission In-plane visibility &    0.99 & 1.00 \\
 E$_0$ (eV) &    1.82 & 1.81 \\
$\mu_{e-h}^2$ (Debye$^2$)&    $2.57 \times 10^4$ & $9.46 \times 10^4$ \\
 $\Gamma_R$ (1/s) &  $4.74 \times 10^2$ & $1.70\times 10^6$ \\
 $\tau_R$ (ns) &  $2.11 \times 10^6$ & $5.89 \times 10^5$ \\
 ODMR & unlikely & unlikely \\
\hline
\end{tabular}
\caption{Complete photophysical properties of defects in hBN for quantum technologies.}
  \label{tab:defect_quantum}
\end{table}
\begin{figure*}[ht]
    \centering
    \includegraphics[width = 1\textwidth]{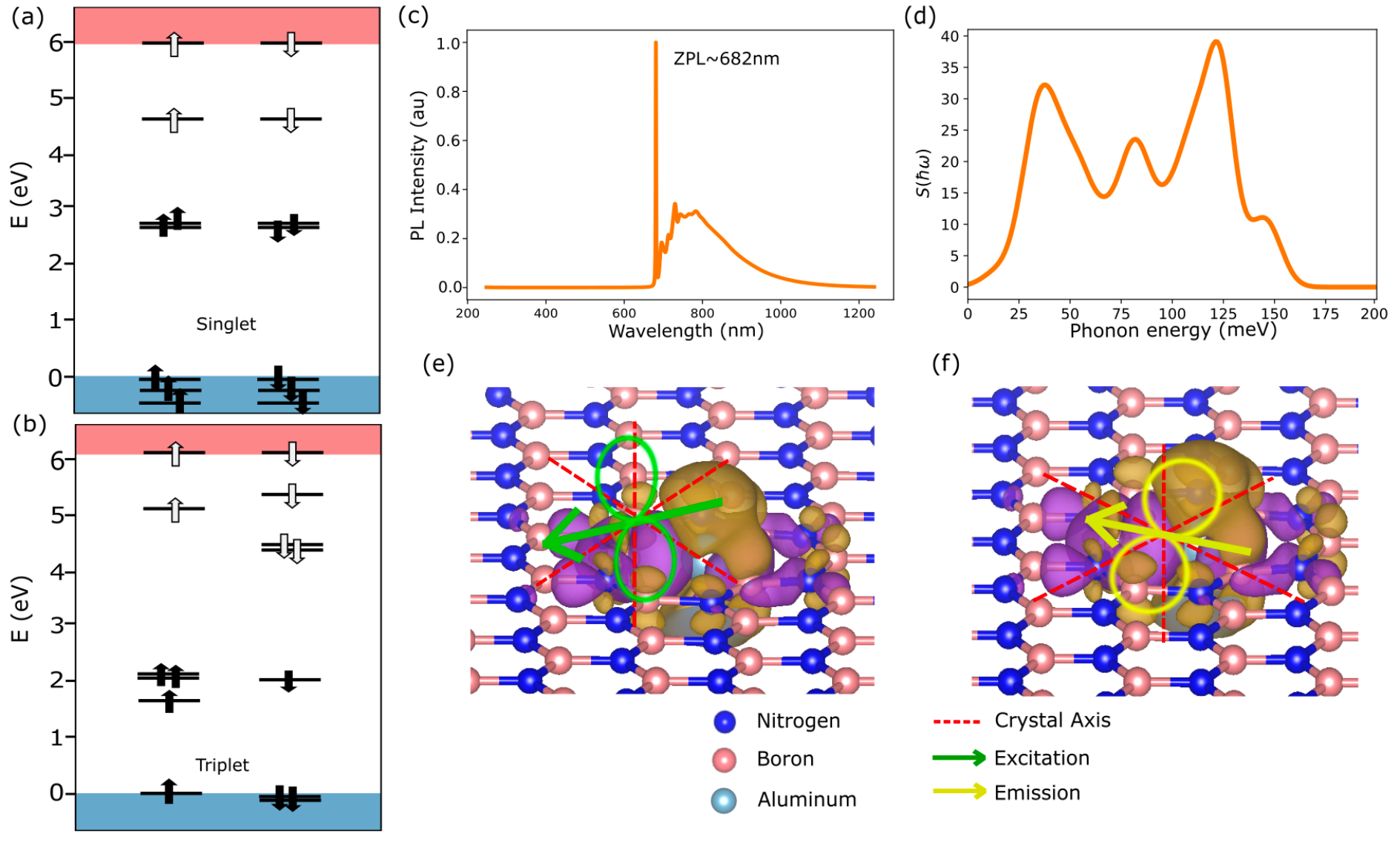}
    \caption{Photophysical properties of Al$_\text{N}$ defect. (a) Single-particle electronic transition of a singlet configuration with the band gap of 6.02 eV whereas (b) a triplet configuration with the band gap of 6.16 eV where the filled and unfilled arrows signify the occupied and unoccupied states. The blue and red shaded areas illustrate the valence and conduction bands, respectively. (c) Calculated photoluminescence spectrum. (d) Spectral function of the phonon energy distribution. (e) The excitation dipole orientation (green arrow) and polarization (dipole pattern). (f) The emission dipole orientation (yellow arrow) and polarization (dipole pattern).}
    \label{fig:AlN}
\end{figure*}
\begin{figure*}[ht]
    \centering
    \includegraphics[width = 1\textwidth]{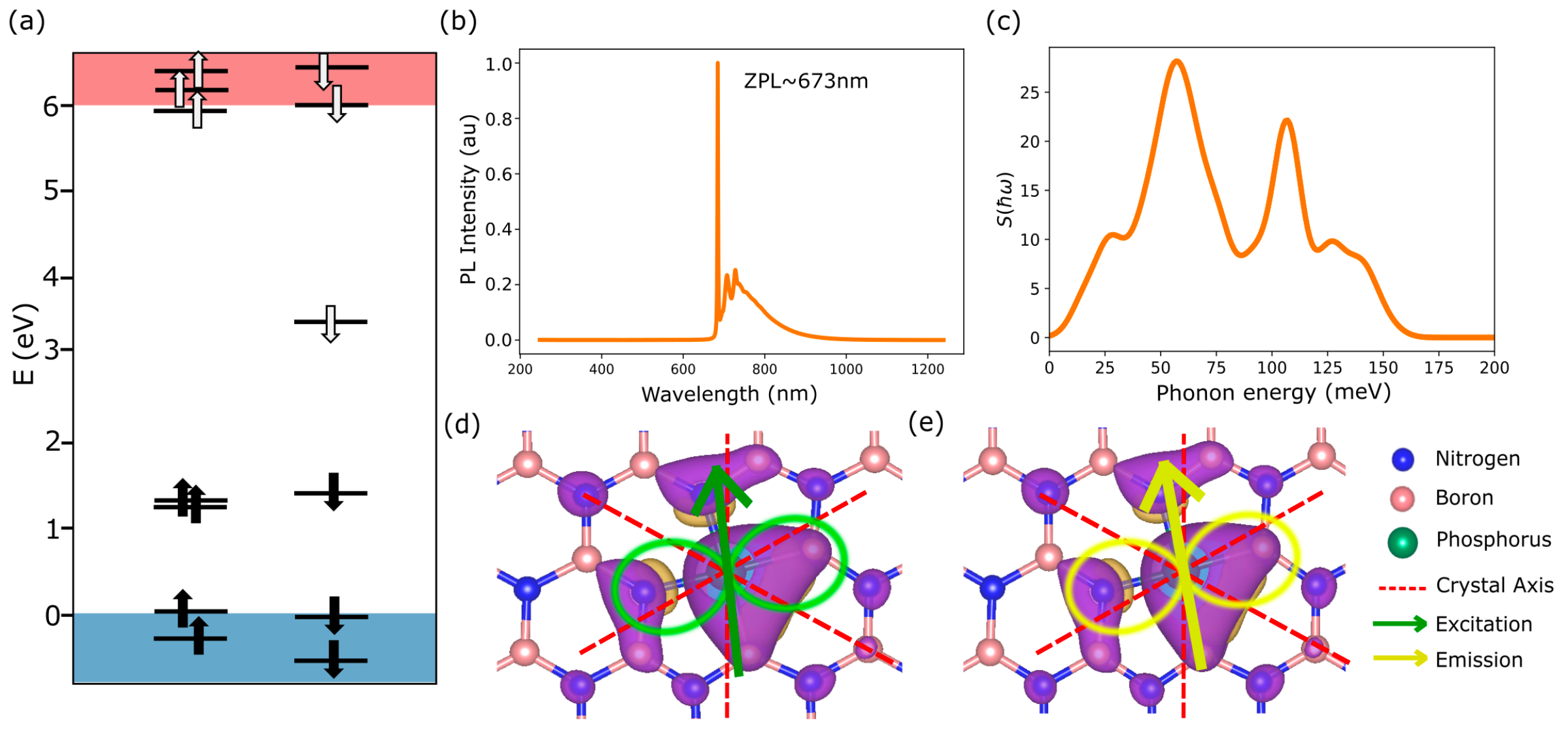}
    \caption{Photophysical properties of P$_\text{N}$V$_\text{B}$ defect. (a) Single-particle electronic transition where the filled and unfilled arrows signify the occupied and unoccupied states. The blue and red shaded areas illustrate the valence and conduction bands, respectively. The band gap is extracted to be 6.02 eV. (b) Calculated photoluminescence spectrum. (c) Spectral function of the phonon energy distribution. (d) The excitation dipole orientation (green arrow) and polarization (dipole pattern). (e) The emission dipole orientation (yellow arrow) and polarization (dipole pattern).}
    \label{fig:PNVB}
\end{figure*}

Their stability was first investigated as shown in the Supplementary Section S3. Under both nitrogen-rich and poor conditions, the neutral charge state of both defects is thermodynamically stable. Turning to explore their electronic properties as listed in Tab.\ \ref{tab:defect_quantum}, both defects inherit compatible emission wavelengths with the NV center (which has its ZPL at 637 nm with very broad PSB) or the hydrogen-$\alpha$ Fraunhofer line (656 nm; would require fine-tuning with strain \cite{Cholsuk2022}). Fig.\ \ref{fig:AlN} depicts photophysical properties of Al$_\text{N}$. Its most preferred configuration is a singlet state as shown in Fig.\ \ref{fig:AlN}(a). However, a triplet configuration can also be formed due to full electron occupation as shown in Fig.\ \ref{fig:AlN}(b) when constraining it as a triplet. This suggests the presence of both triplet and singlet configurations, which might allow a decay pathway between them via the intersystem crossing mechanism. As the singlet configuration is more favorable than the triplet one due to lower total energy, ODMR might not exist in this defect. For quantum sensing (magnetometry), this defect is probably unsuitable. Focusing on the singlet configuration, the ZPL yields 682 nm, which may facilitate coupling between Al$_\text{N}$ and a strong phonon mode of the NV center without the need for frequency conversion. For the PSB as illustrated in Figs. \ref{fig:AlN}(b) and \ref{fig:AlN}(c), it is apparent that PSB of Al$_\text{N}$ is weaker than that of NV center, indicating the low phonon coupling, as also confirmed by the HR factor of approximately 2.94. Next, the lifetime is estimated and found in the order of milliseconds, which might be good for a quantum memory, but not ideal for quantum communication. Finally, we also investigated the dipole orientation as displayed in Figs.\ \ref{fig:AlN}(e)-(f). We found that the excitation polarization has strong out-of-plane contribution which would result in a visibility of only 61\% (aligned with 9.66$^\circ$ from the nearest crystal axis). This would not be an issue, as an inefficient excitation dipole can be accounted for by using a higher excitation laser power. The emission is fully polarized in-plane (i.e., one would observe a near-ideal polarization visibility in reflection) and has an angle of 5.43$^\circ$ away from the crystal axis.\\
\indent As for the P$_\text{N}$V$_\text{B}$ defect, its electronic structure shown in Fig.\ \ref{fig:PNVB}(a) prefers the doublet-spin configuration with its ZPL at 673 nm as the triplet and singlet configurations have partial electron occupation and also a higher total energy than the doublet one. This likely pinpoints the absence of an intersystem-crossing feature in contrast to the NV center; however, its emission wavelength is still close to both NV center and the hydrogen-$\alpha$ Fraunhofer line. For phonon coupling, its HR factor is quite weak at 1.87, implying strong coupling into the ZPL (see Fig.\ \ref{fig:PNVB}(b)-(c)). Furthermore, Tab.\ \ref{tab:defect_quantum} also shows the lifetime of this defect, which is in the order of microseconds. For communication applications, this is not ideal. Lastly, its excitation and emission polarizations are orientated in the in-plane directions, making 5.02$^\circ$ and 16.85$^\circ$ away from the axis, respectively. This high visibility of polarization allows one for easy out-coupling in the typical reflection geometry.
\section{Conclusion}
In conclusion, this work highlights the importance of implementing all optical properties when identifying quantum emitters in solid-state crystals. In particular, measuring the dipole orientations can be achieved with a very simple experiment and easily compared with DFT calculations. It is important that not only a single property is compared but a larger number of photophysical characteristics. This reduces the risk of misassignments. We showcase this by calculating the optical fingerprint of carbon-based defect complexes and compare the results with a previous experiment. It seems that the C$_2$C$_2$ is a promising candidate that could be responsible for one of the 2 eV emitters in hBN. We also calculate the fingerprint of other theoretical defects, the Al$_\text{N}$ and P$_\text{N}$V$_\text{B}$ complexes, for quantum applications and comment on their suitability for these applications. As it turns out, just selecting defects based on their optical transition energy is not sufficient for selecting defects for quantum technologies, as their dipole polarization or lifetime might prevent specific applications. This work therefore provides an important guide for the selection and tailoring of quantum defects for optical quantum technologies.
\newpage
\section*{Data availability}
All data from this work is available from the authors upon reasonable request.

\section*{Notes}
The authors declare no competing financial interest.

\begin{acknowledgments}
This work was funded by the Deutsche Forschungsgemeinschaft (DFG, German Research Foundation) - Projektnummer 445275953. The authors acknowledge support by the German Space Agency DLR with funds provided by the Federal Ministry for Economic Affairs and Climate Action BMWK under grant number 50WM2165 (QUICK3) and 50RP2200 (QuVeKS). T.V. is funded by the Federal Ministry of Education and Research (BMBF) under grant number 13N16292. C.C. is grateful to the Development and Promotion of Science and Technology Talents Project (DPST) scholarship by the Royal Thai Government. S.S. acknowledges funding support by Mahidol University (Fundamental Fund: fiscal year 2023 by National Science Research and Innovation Fund (NSRF)) and from the NSRF via the Program Management Unit for Human Resources \& Institutional Development, Research and Innovation (grant number B05F650024). The computational experiments were supported by resources of the Friedrich Schiller University Jena supported in part by DFG grants INST 275/334-1 FUGG and INST 275/363-1 FUGG. We are grateful to Joel Davidsson for the source code of transition dipole moments for two wavefunctions.
\end{acknowledgments}


\bibliography{main}

\providecommand{\noopsort}[1]{}\providecommand{\singleletter}[1]{#1}
\begin{thebibliography}{58}%
\makeatletter
\providecommand \@ifxundefined [1]{%
 \@ifx{#1\undefined}
}%
\providecommand \@ifnum [1]{%
 \ifnum #1\expandafter \@firstoftwo
 \else \expandafter \@secondoftwo
 \fi
}%
\providecommand \@ifx [1]{%
 \ifx #1\expandafter \@firstoftwo
 \else \expandafter \@secondoftwo
 \fi
}%
\providecommand \natexlab [1]{#1}%
\providecommand \enquote  [1]{``#1''}%
\providecommand \bibnamefont  [1]{#1}%
\providecommand \bibfnamefont [1]{#1}%
\providecommand \citenamefont [1]{#1}%
\providecommand \href@noop [0]{\@secondoftwo}%
\providecommand \href [0]{\begingroup \@sanitize@url \@href}%
\providecommand \@href[1]{\@@startlink{#1}\@@href}%
\providecommand \@@href[1]{\endgroup#1\@@endlink}%
\providecommand \@sanitize@url [0]{\catcode `\\12\catcode `\$12\catcode
  `\&12\catcode `\#12\catcode `\^12\catcode `\_12\catcode `\%12\relax}%
\providecommand \@@startlink[1]{}%
\providecommand \@@endlink[0]{}%
\providecommand \url  [0]{\begingroup\@sanitize@url \@url }%
\providecommand \@url [1]{\endgroup\@href {#1}{\urlprefix }}%
\providecommand \urlprefix  [0]{URL }%
\providecommand \Eprint [0]{\href }%
\providecommand \doibase [0]{https://doi.org/}%
\providecommand \selectlanguage [0]{\@gobble}%
\providecommand \bibinfo  [0]{\@secondoftwo}%
\providecommand \bibfield  [0]{\@secondoftwo}%
\providecommand \translation [1]{[#1]}%
\providecommand \BibitemOpen [0]{}%
\providecommand \bibitemStop [0]{}%
\providecommand \bibitemNoStop [0]{.\EOS\space}%
\providecommand \EOS [0]{\spacefactor3000\relax}%
\providecommand \BibitemShut  [1]{\csname bibitem#1\endcsname}%
\let\auto@bib@innerbib\@empty
\bibitem [{\citenamefont {Bradac}\ \emph {et~al.}(2019)\citenamefont {Bradac},
  \citenamefont {Gao}, \citenamefont {Forneris}, \citenamefont {Trusheim},\
  and\ \citenamefont {Aharonovich}}]{Bradac2019}%
  \BibitemOpen
  \bibfield  {author} {\bibinfo {author} {\bibfnamefont {C.}~\bibnamefont
  {Bradac}}, \bibinfo {author} {\bibfnamefont {W.}~\bibnamefont {Gao}},
  \bibinfo {author} {\bibfnamefont {J.}~\bibnamefont {Forneris}}, \bibinfo
  {author} {\bibfnamefont {M.~E.}\ \bibnamefont {Trusheim}},\ and\ \bibinfo
  {author} {\bibfnamefont {I.}~\bibnamefont {Aharonovich}},\ }\bibfield
  {title} {\bibinfo {title} {Quantum nanophotonics with group iv defects in
  diamond},\ }\href {https://doi.org/10.1038/s41467-019-13332-w} {\bibfield
  {journal} {\bibinfo  {journal} {Nat. Commun.}\ }\textbf {\bibinfo {volume}
  {10}},\ \bibinfo {pages} {5625} (\bibinfo {year} {2019})}\BibitemShut
  {NoStop}%
\bibitem [{\citenamefont {Senichev}\ \emph {et~al.}(2021)\citenamefont
  {Senichev}, \citenamefont {Martin}, \citenamefont {Peana}, \citenamefont
  {Sychev}, \citenamefont {Xu}, \citenamefont {Lagutchev}, \citenamefont
  {Boltasseva},\ and\ \citenamefont {Shalaev}}]{doi:10.1126/sciadv.abj0627}%
  \BibitemOpen
  \bibfield  {author} {\bibinfo {author} {\bibfnamefont {A.}~\bibnamefont
  {Senichev}}, \bibinfo {author} {\bibfnamefont {Z.~O.}\ \bibnamefont
  {Martin}}, \bibinfo {author} {\bibfnamefont {S.}~\bibnamefont {Peana}},
  \bibinfo {author} {\bibfnamefont {D.}~\bibnamefont {Sychev}}, \bibinfo
  {author} {\bibfnamefont {X.}~\bibnamefont {Xu}}, \bibinfo {author}
  {\bibfnamefont {A.~S.}\ \bibnamefont {Lagutchev}}, \bibinfo {author}
  {\bibfnamefont {A.}~\bibnamefont {Boltasseva}},\ and\ \bibinfo {author}
  {\bibfnamefont {V.~M.}\ \bibnamefont {Shalaev}},\ }\bibfield  {title}
  {\bibinfo {title} {Room-temperature single-photon emitters in silicon
  nitride},\ }\href {https://doi.org/10.1126/sciadv.abj0627} {\bibfield
  {journal} {\bibinfo  {journal} {Sci. Adv.}\ }\textbf {\bibinfo {volume}
  {7}},\ \bibinfo {pages} {1} (\bibinfo {year} {2021})}\BibitemShut {NoStop}%
\bibitem [{\citenamefont {Vogl}\ \emph
  {et~al.}(2019{\natexlab{a}})\citenamefont {Vogl}, \citenamefont {Doherty},
  \citenamefont {Buchler}, \citenamefont {Lu},\ and\ \citenamefont
  {Lam}}]{Vogl2019}%
  \BibitemOpen
  \bibfield  {author} {\bibinfo {author} {\bibfnamefont {T.}~\bibnamefont
  {Vogl}}, \bibinfo {author} {\bibfnamefont {M.~W.}\ \bibnamefont {Doherty}},
  \bibinfo {author} {\bibfnamefont {B.~C.}\ \bibnamefont {Buchler}}, \bibinfo
  {author} {\bibfnamefont {Y.}~\bibnamefont {Lu}},\ and\ \bibinfo {author}
  {\bibfnamefont {P.~K.}\ \bibnamefont {Lam}},\ }\bibfield  {title} {\bibinfo
  {title} {Atomic localization of quantum emitters in multilayer hexagonal
  boron nitride},\ }\href {https://doi.org/10.1039/c9nr04269e} {\bibfield
  {journal} {\bibinfo  {journal} {Nanoscale}\ }\textbf {\bibinfo {volume}
  {11}},\ \bibinfo {pages} {14362} (\bibinfo {year}
  {2019}{\natexlab{a}})}\BibitemShut {NoStop}%
\bibitem [{\citenamefont {He}\ \emph {et~al.}(2015)\citenamefont {He},
  \citenamefont {Clark}, \citenamefont {Schaibley}, \citenamefont {He},
  \citenamefont {Chen}, \citenamefont {Wei}, \citenamefont {Ding},
  \citenamefont {Zhang}, \citenamefont {Yao}, \citenamefont {Xu},\ and\
  \citenamefont {et~al.}}]{spe-tmd}%
  \BibitemOpen
  \bibfield  {author} {\bibinfo {author} {\bibfnamefont {Y.-M.}\ \bibnamefont
  {He}}, \bibinfo {author} {\bibfnamefont {G.}~\bibnamefont {Clark}}, \bibinfo
  {author} {\bibfnamefont {J.~R.}\ \bibnamefont {Schaibley}}, \bibinfo {author}
  {\bibfnamefont {Y.}~\bibnamefont {He}}, \bibinfo {author} {\bibfnamefont
  {M.-C.}\ \bibnamefont {Chen}}, \bibinfo {author} {\bibfnamefont {Y.-J.}\
  \bibnamefont {Wei}}, \bibinfo {author} {\bibfnamefont {X.}~\bibnamefont
  {Ding}}, \bibinfo {author} {\bibfnamefont {Q.}~\bibnamefont {Zhang}},
  \bibinfo {author} {\bibfnamefont {W.}~\bibnamefont {Yao}}, \bibinfo {author}
  {\bibfnamefont {X.}~\bibnamefont {Xu}},\ and\ \bibinfo {author} {\bibnamefont
  {et~al.}},\ }\bibfield  {title} {\bibinfo {title} {Single quantum emitters in
  monolayer semiconductors},\ }\href {https://doi.org/10.1038/nnano.2015.75}
  {\bibfield  {journal} {\bibinfo  {journal} {Nat. Nanotechnol.}\ }\textbf
  {\bibinfo {volume} {10}},\ \bibinfo {pages} {497} (\bibinfo {year}
  {2015})}\BibitemShut {NoStop}%
\bibitem [{\citenamefont {Srivastava}\ \emph {et~al.}(2015)\citenamefont
  {Srivastava}, \citenamefont {Sidler}, \citenamefont {Allain}, \citenamefont
  {Lembke}, \citenamefont {Kis},\ and\ \citenamefont
  {Imamoğlu}}]{srivastava_sidler_allain_lembke_kis_imamoglu_2015}%
  \BibitemOpen
  \bibfield  {author} {\bibinfo {author} {\bibfnamefont {A.}~\bibnamefont
  {Srivastava}}, \bibinfo {author} {\bibfnamefont {M.}~\bibnamefont {Sidler}},
  \bibinfo {author} {\bibfnamefont {A.~V.}\ \bibnamefont {Allain}}, \bibinfo
  {author} {\bibfnamefont {D.~S.}\ \bibnamefont {Lembke}}, \bibinfo {author}
  {\bibfnamefont {A.}~\bibnamefont {Kis}},\ and\ \bibinfo {author}
  {\bibfnamefont {A.}~\bibnamefont {Imamoğlu}},\ }\bibfield  {title} {\bibinfo
  {title} {Optically active quantum dots in monolayer wse$_2$},\ }\href
  {https://doi.org/10.1038/nnano.2015.60} {\bibfield  {journal} {\bibinfo
  {journal} {Nat. Nanotechnol.}\ }\textbf {\bibinfo {volume} {10}},\ \bibinfo
  {pages} {491} (\bibinfo {year} {2015})}\BibitemShut {NoStop}%
\bibitem [{\citenamefont {Tran}\ \emph {et~al.}(2016)\citenamefont {Tran},
  \citenamefont {Bray}, \citenamefont {Ford}, \citenamefont {Toth},\ and\
  \citenamefont {Aharonovich}}]{Tran2016}%
  \BibitemOpen
  \bibfield  {author} {\bibinfo {author} {\bibfnamefont {T.~T.}\ \bibnamefont
  {Tran}}, \bibinfo {author} {\bibfnamefont {K.}~\bibnamefont {Bray}}, \bibinfo
  {author} {\bibfnamefont {M.~J.}\ \bibnamefont {Ford}}, \bibinfo {author}
  {\bibfnamefont {M.}~\bibnamefont {Toth}},\ and\ \bibinfo {author}
  {\bibfnamefont {I.}~\bibnamefont {Aharonovich}},\ }\bibfield  {title}
  {\bibinfo {title} {Quantum emission from hexagonal boron nitride
  monolayers},\ }\href {https://doi.org/10.1038/nnano.2015.242} {\bibfield
  {journal} {\bibinfo  {journal} {Nat. Nanotechnol.}\ }\textbf {\bibinfo
  {volume} {11}},\ \bibinfo {pages} {37} (\bibinfo {year} {2016})}\BibitemShut
  {NoStop}%
\bibitem [{\citenamefont {Vogl}\ \emph {et~al.}(2018)\citenamefont {Vogl},
  \citenamefont {Campbell}, \citenamefont {Buchler}, \citenamefont {Lu},\ and\
  \citenamefont {Lam}}]{vogl2018}%
  \BibitemOpen
  \bibfield  {author} {\bibinfo {author} {\bibfnamefont {T.}~\bibnamefont
  {Vogl}}, \bibinfo {author} {\bibfnamefont {G.}~\bibnamefont {Campbell}},
  \bibinfo {author} {\bibfnamefont {B.~C.}\ \bibnamefont {Buchler}}, \bibinfo
  {author} {\bibfnamefont {Y.}~\bibnamefont {Lu}},\ and\ \bibinfo {author}
  {\bibfnamefont {P.~K.}\ \bibnamefont {Lam}},\ }\bibfield  {title} {\bibinfo
  {title} {Fabrication and deterministic transfer of high-quality quantum
  emitters in hexagonal boron nitride},\ }\href
  {https://doi.org/10.1021/acsphotonics.8b00127} {\bibfield  {journal}
  {\bibinfo  {journal} {ACS Photonics}\ }\textbf {\bibinfo {volume} {5}},\
  \bibinfo {pages} {2305} (\bibinfo {year} {2018})}\BibitemShut {NoStop}%
\bibitem [{\citenamefont {O'Brien}(2007)}]{Brien}%
  \BibitemOpen
  \bibfield  {author} {\bibinfo {author} {\bibfnamefont {J.~L.}\ \bibnamefont
  {O'Brien}},\ }\bibfield  {title} {\bibinfo {title} {Optical quantum
  computing},\ }\href {https://doi.org/10.1126/science.1142892} {\bibfield
  {journal} {\bibinfo  {journal} {Science}\ }\textbf {\bibinfo {volume}
  {318}},\ \bibinfo {pages} {1567} (\bibinfo {year} {2007})}\BibitemShut
  {NoStop}%
\bibitem [{\citenamefont {Reilly}(2015)}]{Reilly2015}%
  \BibitemOpen
  \bibfield  {author} {\bibinfo {author} {\bibfnamefont {D.~J.}\ \bibnamefont
  {Reilly}},\ }\bibfield  {title} {\bibinfo {title} {Engineering the
  quantum-classical interface of solid-state qubits},\ }\href
  {https://doi.org/10.1038/npjqi.2015.11} {\bibfield  {journal} {\bibinfo
  {journal} {NPJ Quantum Inf.}\ }\textbf {\bibinfo {volume} {1}},\ \bibinfo
  {pages} {15011} (\bibinfo {year} {2015})}\BibitemShut {NoStop}%
\bibitem [{\citenamefont {Pezzagna}\ and\ \citenamefont
  {Meijer}(2021)}]{Pezzagna2021}%
  \BibitemOpen
  \bibfield  {author} {\bibinfo {author} {\bibfnamefont {S.}~\bibnamefont
  {Pezzagna}}\ and\ \bibinfo {author} {\bibfnamefont {J.}~\bibnamefont
  {Meijer}},\ }\bibfield  {title} {\bibinfo {title} {Quantum computer based on
  color centers in diamond},\ }\href {https://doi.org/10.1063/5.0007444}
  {\bibfield  {journal} {\bibinfo  {journal} {Appl. Phys. Rev.}\ }\textbf
  {\bibinfo {volume} {8}},\ \bibinfo {pages} {011308} (\bibinfo {year}
  {2021})}\BibitemShut {NoStop}%
\bibitem [{\citenamefont {Conlon}\ \emph {et~al.}(2023)\citenamefont {Conlon},
  \citenamefont {Vogl}, \citenamefont {Marciniak}, \citenamefont {Pogorelov},
  \citenamefont {Yung}, \citenamefont {Eilenberger}, \citenamefont {Berry},
  \citenamefont {Santana}, \citenamefont {Blatt}, \citenamefont {Monz},
  \citenamefont {Lam},\ and\ \citenamefont {Assad}}]{Conlon2023}%
  \BibitemOpen
  \bibfield  {author} {\bibinfo {author} {\bibfnamefont {L.~O.}\ \bibnamefont
  {Conlon}}, \bibinfo {author} {\bibfnamefont {T.}~\bibnamefont {Vogl}},
  \bibinfo {author} {\bibfnamefont {C.~D.}\ \bibnamefont {Marciniak}}, \bibinfo
  {author} {\bibfnamefont {I.}~\bibnamefont {Pogorelov}}, \bibinfo {author}
  {\bibfnamefont {S.~K.}\ \bibnamefont {Yung}}, \bibinfo {author}
  {\bibfnamefont {F.}~\bibnamefont {Eilenberger}}, \bibinfo {author}
  {\bibfnamefont {D.~W.}\ \bibnamefont {Berry}}, \bibinfo {author}
  {\bibfnamefont {F.~S.}\ \bibnamefont {Santana}}, \bibinfo {author}
  {\bibfnamefont {R.}~\bibnamefont {Blatt}}, \bibinfo {author} {\bibfnamefont
  {T.}~\bibnamefont {Monz}}, \bibinfo {author} {\bibfnamefont {P.~K.}\
  \bibnamefont {Lam}},\ and\ \bibinfo {author} {\bibfnamefont {S.~M.}\
  \bibnamefont {Assad}},\ }\bibfield  {title} {\bibinfo {title} {Approaching
  optimal entangling collective measurements on quantum computing platforms},\
  }\href {https://doi.org/10.1038/s41567-022-01875-7} {\bibfield  {journal}
  {\bibinfo  {journal} {Nat. Phys.}\ }\textbf {\bibinfo {volume} {19}},\
  \bibinfo {pages} {351} (\bibinfo {year} {2023})}\BibitemShut {NoStop}%
\bibitem [{\citenamefont {Chapuran}\ \emph {et~al.}(2009)\citenamefont
  {Chapuran}, \citenamefont {Toliver}, \citenamefont {Peters}, \citenamefont
  {Jackel}, \citenamefont {Goodman}, \citenamefont {Runser}, \citenamefont
  {McNown}, \citenamefont {Dallmann}, \citenamefont {Hughes}, \citenamefont
  {McCabe}, \citenamefont {Nordholt}, \citenamefont {Peterson}, \citenamefont
  {Tyagi}, \citenamefont {Mercer},\ and\ \citenamefont {Dardy}}]{Chapuran2009}%
  \BibitemOpen
  \bibfield  {author} {\bibinfo {author} {\bibfnamefont {T.~E.}\ \bibnamefont
  {Chapuran}}, \bibinfo {author} {\bibfnamefont {P.}~\bibnamefont {Toliver}},
  \bibinfo {author} {\bibfnamefont {N.~A.}\ \bibnamefont {Peters}}, \bibinfo
  {author} {\bibfnamefont {J.}~\bibnamefont {Jackel}}, \bibinfo {author}
  {\bibfnamefont {M.~S.}\ \bibnamefont {Goodman}}, \bibinfo {author}
  {\bibfnamefont {R.~J.}\ \bibnamefont {Runser}}, \bibinfo {author}
  {\bibfnamefont {S.~R.}\ \bibnamefont {McNown}}, \bibinfo {author}
  {\bibfnamefont {N.}~\bibnamefont {Dallmann}}, \bibinfo {author}
  {\bibfnamefont {R.~J.}\ \bibnamefont {Hughes}}, \bibinfo {author}
  {\bibfnamefont {K.~P.}\ \bibnamefont {McCabe}}, \bibinfo {author}
  {\bibfnamefont {J.~E.}\ \bibnamefont {Nordholt}}, \bibinfo {author}
  {\bibfnamefont {C.~G.}\ \bibnamefont {Peterson}}, \bibinfo {author}
  {\bibfnamefont {K.~T.}\ \bibnamefont {Tyagi}}, \bibinfo {author}
  {\bibfnamefont {L.}~\bibnamefont {Mercer}},\ and\ \bibinfo {author}
  {\bibfnamefont {H.}~\bibnamefont {Dardy}},\ }\bibfield  {title} {\bibinfo
  {title} {Optical networking for quantum key distribution and quantum
  communications},\ }\href {https://doi.org/10.1088/1367-2630/11/10/105001}
  {\bibfield  {journal} {\bibinfo  {journal} {New J. Phys.}\ }\textbf {\bibinfo
  {volume} {11}},\ \bibinfo {pages} {105001} (\bibinfo {year}
  {2009})}\BibitemShut {NoStop}%
\bibitem [{\citenamefont {Abasifard}\ \emph {et~al.}(2023)\citenamefont
  {Abasifard}, \citenamefont {Cholsuk}, \citenamefont {Pousa}, \citenamefont
  {Kumar}, \citenamefont {Zand}, \citenamefont {Riel}, \citenamefont {Oi},\
  and\ \citenamefont {Vogl}}]{mostafa-qkd}%
  \BibitemOpen
  \bibfield  {author} {\bibinfo {author} {\bibfnamefont {M.}~\bibnamefont
  {Abasifard}}, \bibinfo {author} {\bibfnamefont {C.}~\bibnamefont {Cholsuk}},
  \bibinfo {author} {\bibfnamefont {R.~G.}\ \bibnamefont {Pousa}}, \bibinfo
  {author} {\bibfnamefont {A.}~\bibnamefont {Kumar}}, \bibinfo {author}
  {\bibfnamefont {A.}~\bibnamefont {Zand}}, \bibinfo {author} {\bibfnamefont
  {T.}~\bibnamefont {Riel}}, \bibinfo {author} {\bibfnamefont {D.~K.~L.}\
  \bibnamefont {Oi}},\ and\ \bibinfo {author} {\bibfnamefont {T.}~\bibnamefont
  {Vogl}},\ }\bibfield  {title} {\bibinfo {title} {The ideal wavelength for
  daylight free-space quantum key distribution},\ }\href@noop {} {\bibfield
  {journal} {\bibinfo  {journal} {arXiv.org, e-Print Arch., Phys.}\ ,\ \bibinfo
  {pages} {arXiv:2303.02106}} (\bibinfo {year} {2023})},\ \Eprint
  {https://arxiv.org/abs/2303.02106} {arXiv:2303.02106 [quant-ph]} \BibitemShut
  {NoStop}%
\bibitem [{\citenamefont {Aslam}\ \emph {et~al.}(2023)\citenamefont {Aslam},
  \citenamefont {Zhou}, \citenamefont {Urbach}, \citenamefont {Turner},
  \citenamefont {Walsworth}, \citenamefont {Lukin},\ and\ \citenamefont
  {Park}}]{Aslam2023}%
  \BibitemOpen
  \bibfield  {author} {\bibinfo {author} {\bibfnamefont {N.}~\bibnamefont
  {Aslam}}, \bibinfo {author} {\bibfnamefont {H.}~\bibnamefont {Zhou}},
  \bibinfo {author} {\bibfnamefont {E.~K.}\ \bibnamefont {Urbach}}, \bibinfo
  {author} {\bibfnamefont {M.~J.}\ \bibnamefont {Turner}}, \bibinfo {author}
  {\bibfnamefont {R.~L.}\ \bibnamefont {Walsworth}}, \bibinfo {author}
  {\bibfnamefont {M.~D.}\ \bibnamefont {Lukin}},\ and\ \bibinfo {author}
  {\bibfnamefont {H.}~\bibnamefont {Park}},\ }\bibfield  {title} {\bibinfo
  {title} {Quantum sensors for biomedical applications},\ }\href
  {https://doi.org/10.1038/s42254-023-00558-3} {\bibfield  {journal} {\bibinfo
  {journal} {Nat. Rev. Phys.}\ }\textbf {\bibinfo {volume} {5}},\ \bibinfo
  {pages} {157} (\bibinfo {year} {2023})}\BibitemShut {NoStop}%
\bibitem [{\citenamefont {Ahmadi}\ \emph {et~al.}(2023)\citenamefont {Ahmadi},
  \citenamefont {Schwertfeger}, \citenamefont {Werner}, \citenamefont {Wiese},
  \citenamefont {Lester}, \citenamefont {Ros}, \citenamefont {Krause},
  \citenamefont {Ritter}, \citenamefont {Abasifard}, \citenamefont {Cholsuk},
  \citenamefont {Krämer}, \citenamefont {Atzeni}, \citenamefont {Gündoğan},
  \citenamefont {Sachidananda}, \citenamefont {Pardo}, \citenamefont {Nolte},
  \citenamefont {Lohrmann}, \citenamefont {Ling}, \citenamefont
  {Bartholomäus}, \citenamefont {Corrielli}, \citenamefont {Krutzik},\ and\
  \citenamefont {Vogl}}]{quick3}%
  \BibitemOpen
  \bibfield  {author} {\bibinfo {author} {\bibfnamefont {N.}~\bibnamefont
  {Ahmadi}}, \bibinfo {author} {\bibfnamefont {S.}~\bibnamefont
  {Schwertfeger}}, \bibinfo {author} {\bibfnamefont {P.}~\bibnamefont
  {Werner}}, \bibinfo {author} {\bibfnamefont {L.}~\bibnamefont {Wiese}},
  \bibinfo {author} {\bibfnamefont {J.}~\bibnamefont {Lester}}, \bibinfo
  {author} {\bibfnamefont {E.~D.}\ \bibnamefont {Ros}}, \bibinfo {author}
  {\bibfnamefont {J.}~\bibnamefont {Krause}}, \bibinfo {author} {\bibfnamefont
  {S.}~\bibnamefont {Ritter}}, \bibinfo {author} {\bibfnamefont
  {M.}~\bibnamefont {Abasifard}}, \bibinfo {author} {\bibfnamefont
  {C.}~\bibnamefont {Cholsuk}}, \bibinfo {author} {\bibfnamefont {R.~G.}\
  \bibnamefont {Krämer}}, \bibinfo {author} {\bibfnamefont {S.}~\bibnamefont
  {Atzeni}}, \bibinfo {author} {\bibfnamefont {M.}~\bibnamefont {Gündoğan}},
  \bibinfo {author} {\bibfnamefont {S.}~\bibnamefont {Sachidananda}}, \bibinfo
  {author} {\bibfnamefont {D.}~\bibnamefont {Pardo}}, \bibinfo {author}
  {\bibfnamefont {S.}~\bibnamefont {Nolte}}, \bibinfo {author} {\bibfnamefont
  {A.}~\bibnamefont {Lohrmann}}, \bibinfo {author} {\bibfnamefont
  {A.}~\bibnamefont {Ling}}, \bibinfo {author} {\bibfnamefont {J.}~\bibnamefont
  {Bartholomäus}}, \bibinfo {author} {\bibfnamefont {G.}~\bibnamefont
  {Corrielli}}, \bibinfo {author} {\bibfnamefont {M.}~\bibnamefont {Krutzik}},\
  and\ \bibinfo {author} {\bibfnamefont {T.}~\bibnamefont {Vogl}},\ }\bibfield
  {title} {\bibinfo {title} {Quick$^3$ -- design of a satellite-based quantum
  light source for quantum communication and extended physical theory tests in
  space},\ }\href@noop {} {\bibfield  {journal} {\bibinfo  {journal}
  {arXiv.org, e-Print Arch., Phys.}\ ,\ \bibinfo {pages} {arXiv:2301.11177}}
  (\bibinfo {year} {2023})},\ \Eprint {https://arxiv.org/abs/2301.11177}
  {arXiv:2301.11177 [quant-ph]} \BibitemShut {NoStop}%
\bibitem [{\citenamefont {Lvovsky}\ \emph {et~al.}(2009)\citenamefont
  {Lvovsky}, \citenamefont {Sanders},\ and\ \citenamefont
  {Tittel}}]{Lvovsky2009}%
  \BibitemOpen
  \bibfield  {author} {\bibinfo {author} {\bibfnamefont {A.~I.}\ \bibnamefont
  {Lvovsky}}, \bibinfo {author} {\bibfnamefont {B.~C.}\ \bibnamefont
  {Sanders}},\ and\ \bibinfo {author} {\bibfnamefont {W.}~\bibnamefont
  {Tittel}},\ }\bibfield  {title} {\bibinfo {title} {Optical quantum memory},\
  }\href {https://doi.org/10.1038/nphoton.2009.231} {\bibfield  {journal}
  {\bibinfo  {journal} {Nat. Photonics}\ }\textbf {\bibinfo {volume} {3}},\
  \bibinfo {pages} {706} (\bibinfo {year} {2009})}\BibitemShut {NoStop}%
\bibitem [{\citenamefont {Pingault}\ \emph {et~al.}(2014)\citenamefont
  {Pingault}, \citenamefont {Becker}, \citenamefont {Schulte}, \citenamefont
  {Arend}, \citenamefont {Hepp}, \citenamefont {Godde}, \citenamefont
  {Tartakovskii}, \citenamefont {Markham}, \citenamefont {Becher},\ and\
  \citenamefont {Atatüre}}]{Pingault2014}%
  \BibitemOpen
  \bibfield  {author} {\bibinfo {author} {\bibfnamefont {B.}~\bibnamefont
  {Pingault}}, \bibinfo {author} {\bibfnamefont {J.~N.}\ \bibnamefont
  {Becker}}, \bibinfo {author} {\bibfnamefont {C.~H.}\ \bibnamefont {Schulte}},
  \bibinfo {author} {\bibfnamefont {C.}~\bibnamefont {Arend}}, \bibinfo
  {author} {\bibfnamefont {C.}~\bibnamefont {Hepp}}, \bibinfo {author}
  {\bibfnamefont {T.}~\bibnamefont {Godde}}, \bibinfo {author} {\bibfnamefont
  {A.~I.}\ \bibnamefont {Tartakovskii}}, \bibinfo {author} {\bibfnamefont
  {M.}~\bibnamefont {Markham}}, \bibinfo {author} {\bibfnamefont
  {C.}~\bibnamefont {Becher}},\ and\ \bibinfo {author} {\bibfnamefont
  {M.}~\bibnamefont {Atatüre}},\ }\bibfield  {title} {\bibinfo {title}
  {All-optical formation of coherent dark states of silicon-vacancy spins in
  diamond},\ }\href {https://doi.org/10.1103/PhysRevLett.113.263601} {\bibfield
   {journal} {\bibinfo  {journal} {Phys. Rev. Lett.}\ }\textbf {\bibinfo
  {volume} {113}},\ \bibinfo {pages} {263601} (\bibinfo {year}
  {2014})}\BibitemShut {NoStop}%
\bibitem [{\citenamefont {Liu}\ \emph {et~al.}(2020)\citenamefont {Liu},
  \citenamefont {Wang}, \citenamefont {Li}, \citenamefont {Yu}, \citenamefont
  {Ke}, \citenamefont {Meng}, \citenamefont {Tang}, \citenamefont {Li},\ and\
  \citenamefont {Guo}}]{LIU2020114251}%
  \BibitemOpen
  \bibfield  {author} {\bibinfo {author} {\bibfnamefont {W.}~\bibnamefont
  {Liu}}, \bibinfo {author} {\bibfnamefont {Y.-T.}\ \bibnamefont {Wang}},
  \bibinfo {author} {\bibfnamefont {Z.-P.}\ \bibnamefont {Li}}, \bibinfo
  {author} {\bibfnamefont {S.}~\bibnamefont {Yu}}, \bibinfo {author}
  {\bibfnamefont {Z.-J.}\ \bibnamefont {Ke}}, \bibinfo {author} {\bibfnamefont
  {Y.}~\bibnamefont {Meng}}, \bibinfo {author} {\bibfnamefont {J.-S.}\
  \bibnamefont {Tang}}, \bibinfo {author} {\bibfnamefont {C.-F.}\ \bibnamefont
  {Li}},\ and\ \bibinfo {author} {\bibfnamefont {G.-C.}\ \bibnamefont {Guo}},\
  }\bibfield  {title} {\bibinfo {title} {An ultrastable and robust
  single-photon emitter in hexagonal boron nitride},\ }\href
  {https://doi.org/https://doi.org/10.1016/j.physe.2020.114251} {\bibfield
  {journal} {\bibinfo  {journal} {Physica E Low Dimens. Syst. Nanostruct.}\
  }\textbf {\bibinfo {volume} {124}},\ \bibinfo {pages} {114251} (\bibinfo
  {year} {2020})}\BibitemShut {NoStop}%
\bibitem [{\citenamefont {Vogl}\ \emph
  {et~al.}(2019{\natexlab{b}})\citenamefont {Vogl}, \citenamefont {Sripathy},
  \citenamefont {Sharma}, \citenamefont {Reddy}, \citenamefont {Sullivan},
  \citenamefont {Machacek}, \citenamefont {Zhang}, \citenamefont {Karouta},
  \citenamefont {Buchler}, \citenamefont {Doherty}, \citenamefont {Lu},\ and\
  \citenamefont {Lam}}]{Vogl2019-radiation}%
  \BibitemOpen
  \bibfield  {author} {\bibinfo {author} {\bibfnamefont {T.}~\bibnamefont
  {Vogl}}, \bibinfo {author} {\bibfnamefont {K.}~\bibnamefont {Sripathy}},
  \bibinfo {author} {\bibfnamefont {A.}~\bibnamefont {Sharma}}, \bibinfo
  {author} {\bibfnamefont {P.}~\bibnamefont {Reddy}}, \bibinfo {author}
  {\bibfnamefont {J.}~\bibnamefont {Sullivan}}, \bibinfo {author}
  {\bibfnamefont {J.~R.}\ \bibnamefont {Machacek}}, \bibinfo {author}
  {\bibfnamefont {L.}~\bibnamefont {Zhang}}, \bibinfo {author} {\bibfnamefont
  {F.}~\bibnamefont {Karouta}}, \bibinfo {author} {\bibfnamefont {B.~C.}\
  \bibnamefont {Buchler}}, \bibinfo {author} {\bibfnamefont {M.~W.}\
  \bibnamefont {Doherty}}, \bibinfo {author} {\bibfnamefont {Y.}~\bibnamefont
  {Lu}},\ and\ \bibinfo {author} {\bibfnamefont {P.~K.}\ \bibnamefont {Lam}},\
  }\bibfield  {title} {\bibinfo {title} {Radiation tolerance of two-dimensional
  material-based devices for space applications},\ }\href
  {https://doi.org/10.1038/s41467-019-09219-5} {\bibfield  {journal} {\bibinfo
  {journal} {Nat. Commun.}\ }\textbf {\bibinfo {volume} {10}},\ \bibinfo
  {pages} {1202} (\bibinfo {year} {2019}{\natexlab{b}})}\BibitemShut {NoStop}%
\bibitem [{\citenamefont {Kianinia}\ \emph {et~al.}(2017)\citenamefont
  {Kianinia}, \citenamefont {Regan}, \citenamefont {Tawfik}, \citenamefont
  {Tran}, \citenamefont {Ford}, \citenamefont {Aharonovich},\ and\
  \citenamefont {Toth}}]{Kianinia2017}%
  \BibitemOpen
  \bibfield  {author} {\bibinfo {author} {\bibfnamefont {M.}~\bibnamefont
  {Kianinia}}, \bibinfo {author} {\bibfnamefont {B.}~\bibnamefont {Regan}},
  \bibinfo {author} {\bibfnamefont {S.~A.}\ \bibnamefont {Tawfik}}, \bibinfo
  {author} {\bibfnamefont {T.~T.}\ \bibnamefont {Tran}}, \bibinfo {author}
  {\bibfnamefont {M.~J.}\ \bibnamefont {Ford}}, \bibinfo {author}
  {\bibfnamefont {I.}~\bibnamefont {Aharonovich}},\ and\ \bibinfo {author}
  {\bibfnamefont {M.}~\bibnamefont {Toth}},\ }\bibfield  {title} {\bibinfo
  {title} {Robust solid-state quantum system operating at 800 k},\ }\href
  {https://doi.org/10.1021/acsphotonics.7b00086} {\bibfield  {journal}
  {\bibinfo  {journal} {ACS Photonics}\ }\textbf {\bibinfo {volume} {4}},\
  \bibinfo {pages} {768} (\bibinfo {year} {2017})}\BibitemShut {NoStop}%
\bibitem [{\citenamefont {Kumar}\ \emph
  {et~al.}(2023{\natexlab{a}})\citenamefont {Kumar}, \citenamefont {Çağlar
  Samaner}, \citenamefont {Cholsuk}, \citenamefont {Matthes}, \citenamefont
  {Paçal}, \citenamefont {Oyun}, \citenamefont {Zand}, \citenamefont
  {Chapman}, \citenamefont {Saerens}, \citenamefont {Grange}, \citenamefont
  {Suwanna}, \citenamefont {Ateş},\ and\ \citenamefont {Vogl}}]{anand-dipole}%
  \BibitemOpen
  \bibfield  {author} {\bibinfo {author} {\bibfnamefont {A.}~\bibnamefont
  {Kumar}}, \bibinfo {author} {\bibnamefont {Çağlar Samaner}}, \bibinfo
  {author} {\bibfnamefont {C.}~\bibnamefont {Cholsuk}}, \bibinfo {author}
  {\bibfnamefont {T.}~\bibnamefont {Matthes}}, \bibinfo {author} {\bibfnamefont
  {S.}~\bibnamefont {Paçal}}, \bibinfo {author} {\bibfnamefont
  {Y.}~\bibnamefont {Oyun}}, \bibinfo {author} {\bibfnamefont {A.}~\bibnamefont
  {Zand}}, \bibinfo {author} {\bibfnamefont {R.~J.}\ \bibnamefont {Chapman}},
  \bibinfo {author} {\bibfnamefont {G.}~\bibnamefont {Saerens}}, \bibinfo
  {author} {\bibfnamefont {R.}~\bibnamefont {Grange}}, \bibinfo {author}
  {\bibfnamefont {S.}~\bibnamefont {Suwanna}}, \bibinfo {author} {\bibfnamefont
  {S.}~\bibnamefont {Ateş}},\ and\ \bibinfo {author} {\bibfnamefont
  {T.}~\bibnamefont {Vogl}},\ }\bibfield  {title} {\bibinfo {title}
  {Polarization dynamics of solid-state quantum emitters},\ }\href@noop {}
  {\bibfield  {journal} {\bibinfo  {journal} {arXiv.org, e-Print Arch., Phys.}\
  ,\ \bibinfo {pages} {arXiv:2303.04732}} (\bibinfo {year}
  {2023}{\natexlab{a}})},\ \Eprint {https://arxiv.org/abs/2303.04732}
  {arXiv:2303.04732 [quant-ph]} \BibitemShut {NoStop}%
\bibitem [{\citenamefont {Cholsuk}\ \emph {et~al.}(2022)\citenamefont
  {Cholsuk}, \citenamefont {Suwanna},\ and\ \citenamefont
  {Vogl}}]{Cholsuk2022}%
  \BibitemOpen
  \bibfield  {author} {\bibinfo {author} {\bibfnamefont {C.}~\bibnamefont
  {Cholsuk}}, \bibinfo {author} {\bibfnamefont {S.}~\bibnamefont {Suwanna}},\
  and\ \bibinfo {author} {\bibfnamefont {T.}~\bibnamefont {Vogl}},\ }\bibfield
  {title} {\bibinfo {title} {Tailoring the emission wavelength of color centers
  in hexagonal boron nitride for quantum applications},\ }\href
  {https://doi.org/10.3390/nano12142427} {\bibfield  {journal} {\bibinfo
  {journal} {Nanomaterials}\ }\textbf {\bibinfo {volume} {12}},\ \bibinfo
  {pages} {2427} (\bibinfo {year} {2022})}\BibitemShut {NoStop}%
\bibitem [{\citenamefont {Mackoit-Sinkevičienė}\ \emph
  {et~al.}(2019)\citenamefont {Mackoit-Sinkevičienė}, \citenamefont
  {Maciaszek}, \citenamefont {de~Walle},\ and\ \citenamefont
  {Alkauskas}}]{Cdimer-4eV}%
  \BibitemOpen
  \bibfield  {author} {\bibinfo {author} {\bibfnamefont {M.}~\bibnamefont
  {Mackoit-Sinkevičienė}}, \bibinfo {author} {\bibfnamefont {M.}~\bibnamefont
  {Maciaszek}}, \bibinfo {author} {\bibfnamefont {C.~G.~V.}\ \bibnamefont
  {de~Walle}},\ and\ \bibinfo {author} {\bibfnamefont {A.}~\bibnamefont
  {Alkauskas}},\ }\bibfield  {title} {\bibinfo {title} {Carbon dimer defect as
  a source of the 4.1 ev luminescence in hexagonal boron nitride},\ }\href
  {https://doi.org/10.1063/1.5124153} {\bibfield  {journal} {\bibinfo
  {journal} {Appl. Phys. Lett.}\ }\textbf {\bibinfo {volume} {115}},\ \bibinfo
  {pages} {212101} (\bibinfo {year} {2019})}\BibitemShut {NoStop}%
\bibitem [{\citenamefont {Dietrich}\ \emph {et~al.}(2018)\citenamefont
  {Dietrich}, \citenamefont {Bürk}, \citenamefont {Steiger}, \citenamefont
  {Antoniuk}, \citenamefont {Tran}, \citenamefont {Nguyen}, \citenamefont
  {Aharonovich}, \citenamefont {Jelezko},\ and\ \citenamefont
  {Kubanek}}]{Dietrich2018}%
  \BibitemOpen
  \bibfield  {author} {\bibinfo {author} {\bibfnamefont {A.}~\bibnamefont
  {Dietrich}}, \bibinfo {author} {\bibfnamefont {M.}~\bibnamefont {Bürk}},
  \bibinfo {author} {\bibfnamefont {E.~S.}\ \bibnamefont {Steiger}}, \bibinfo
  {author} {\bibfnamefont {L.}~\bibnamefont {Antoniuk}}, \bibinfo {author}
  {\bibfnamefont {T.~T.}\ \bibnamefont {Tran}}, \bibinfo {author}
  {\bibfnamefont {M.}~\bibnamefont {Nguyen}}, \bibinfo {author} {\bibfnamefont
  {I.}~\bibnamefont {Aharonovich}}, \bibinfo {author} {\bibfnamefont
  {F.}~\bibnamefont {Jelezko}},\ and\ \bibinfo {author} {\bibfnamefont
  {A.}~\bibnamefont {Kubanek}},\ }\bibfield  {title} {\bibinfo {title}
  {Observation of fourier transform limited lines in hexagonal boron nitride},\
  }\href {https://doi.org/10.1103/PhysRevB.98.081414} {\bibfield  {journal}
  {\bibinfo  {journal} {Phys. Rev. B.}\ }\textbf {\bibinfo {volume} {98}},\
  \bibinfo {pages} {081414} (\bibinfo {year} {2018})}\BibitemShut {NoStop}%
\bibitem [{\citenamefont {Camphausen}\ \emph {et~al.}(2020)\citenamefont
  {Camphausen}, \citenamefont {Marini}, \citenamefont {Tawfik}, \citenamefont
  {Tran}, \citenamefont {Ford},\ and\ \citenamefont
  {Palomba}}]{Camphausen2020}%
  \BibitemOpen
  \bibfield  {author} {\bibinfo {author} {\bibfnamefont {R.}~\bibnamefont
  {Camphausen}}, \bibinfo {author} {\bibfnamefont {L.}~\bibnamefont {Marini}},
  \bibinfo {author} {\bibfnamefont {S.~A.}\ \bibnamefont {Tawfik}}, \bibinfo
  {author} {\bibfnamefont {T.~T.}\ \bibnamefont {Tran}}, \bibinfo {author}
  {\bibfnamefont {M.~J.}\ \bibnamefont {Ford}},\ and\ \bibinfo {author}
  {\bibfnamefont {S.}~\bibnamefont {Palomba}},\ }\bibfield  {title} {\bibinfo
  {title} {Observation of near-infrared sub-poissonian photon emission in
  hexagonal boron nitride at room temperature},\ }\href
  {https://doi.org/10.1063/5.0008242} {\bibfield  {journal} {\bibinfo
  {journal} {APL Photonics}\ }\textbf {\bibinfo {volume} {5}},\ \bibinfo
  {pages} {076103} (\bibinfo {year} {2020})}\BibitemShut {NoStop}%
\bibitem [{\citenamefont {Grosso}\ \emph {et~al.}(2020)\citenamefont {Grosso},
  \citenamefont {Moon}, \citenamefont {Ciccarino}, \citenamefont {Flick},
  \citenamefont {Mendelson}, \citenamefont {Mennel}, \citenamefont {Toth},
  \citenamefont {Aharonovich}, \citenamefont {Narang},\ and\ \citenamefont
  {Englund}}]{Grosso2020}%
  \BibitemOpen
  \bibfield  {author} {\bibinfo {author} {\bibfnamefont {G.}~\bibnamefont
  {Grosso}}, \bibinfo {author} {\bibfnamefont {H.}~\bibnamefont {Moon}},
  \bibinfo {author} {\bibfnamefont {C.~J.}\ \bibnamefont {Ciccarino}}, \bibinfo
  {author} {\bibfnamefont {J.}~\bibnamefont {Flick}}, \bibinfo {author}
  {\bibfnamefont {N.}~\bibnamefont {Mendelson}}, \bibinfo {author}
  {\bibfnamefont {L.}~\bibnamefont {Mennel}}, \bibinfo {author} {\bibfnamefont
  {M.}~\bibnamefont {Toth}}, \bibinfo {author} {\bibfnamefont {I.}~\bibnamefont
  {Aharonovich}}, \bibinfo {author} {\bibfnamefont {P.}~\bibnamefont
  {Narang}},\ and\ \bibinfo {author} {\bibfnamefont {D.~R.}\ \bibnamefont
  {Englund}},\ }\bibfield  {title} {\bibinfo {title} {Low-temperature
  electron-phonon interaction of quantum emitters in hexagonal boron nitride},\
  }\href {https://doi.org/10.1021/acsphotonics.9b01789} {\bibfield  {journal}
  {\bibinfo  {journal} {ACS Photonics}\ }\textbf {\bibinfo {volume} {7}},\
  \bibinfo {pages} {1410} (\bibinfo {year} {2020})}\BibitemShut {NoStop}%
\bibitem [{\citenamefont {Ivády}\ \emph {et~al.}(2018)\citenamefont {Ivády},
  \citenamefont {Abrikosov},\ and\ \citenamefont {Gali}}]{gali2018}%
  \BibitemOpen
  \bibfield  {author} {\bibinfo {author} {\bibfnamefont {V.}~\bibnamefont
  {Ivády}}, \bibinfo {author} {\bibfnamefont {I.~A.}\ \bibnamefont
  {Abrikosov}},\ and\ \bibinfo {author} {\bibfnamefont {A.}~\bibnamefont
  {Gali}},\ }\bibfield  {title} {\bibinfo {title} {First principles calculation
  of spin-related quantities for point defect qubit research},\ }\href
  {https://doi.org/10.1038/s41524-018-0132-5} {\bibfield  {journal} {\bibinfo
  {journal} {Npj Comput. Mater.}\ }\textbf {\bibinfo {volume} {4}},\ \bibinfo
  {pages} {76} (\bibinfo {year} {2018})}\BibitemShut {NoStop}%
\bibitem [{\citenamefont {Sajid}\ \emph {et~al.}(2020)\citenamefont {Sajid},
  \citenamefont {Reimers}, \citenamefont {Kobayashi},\ and\ \citenamefont
  {Ford}}]{Sajid2020-VNNB}%
  \BibitemOpen
  \bibfield  {author} {\bibinfo {author} {\bibfnamefont {A.}~\bibnamefont
  {Sajid}}, \bibinfo {author} {\bibfnamefont {J.~R.}\ \bibnamefont {Reimers}},
  \bibinfo {author} {\bibfnamefont {R.}~\bibnamefont {Kobayashi}},\ and\
  \bibinfo {author} {\bibfnamefont {M.~J.}\ \bibnamefont {Ford}},\ }\bibfield
  {title} {\bibinfo {title} {Theoretical spectroscopy of the
  v$_{\text{n}}$\text{N}$_{\text{b}}$ defect in hexagonal boron nitride},\
  }\href {https://doi.org/10.1103/PhysRevB.102.144104} {\bibfield  {journal}
  {\bibinfo  {journal} {Phys. Rev. B.}\ }\textbf {\bibinfo {volume} {102}},\
  \bibinfo {pages} {144104} (\bibinfo {year} {2020})}\BibitemShut {NoStop}%
\bibitem [{\citenamefont {Auburger}\ and\ \citenamefont
  {Gali}(2021)}]{Auburger2021}%
  \BibitemOpen
  \bibfield  {author} {\bibinfo {author} {\bibfnamefont {P.}~\bibnamefont
  {Auburger}}\ and\ \bibinfo {author} {\bibfnamefont {A.}~\bibnamefont
  {Gali}},\ }\bibfield  {title} {\bibinfo {title} {Towards ab initio
  identification of paramagnetic substitutional carbon defects in hexagonal
  boron nitride acting as quantum bits},\ }\href
  {https://doi.org/10.1103/PhysRevB.104.075410} {\bibfield  {journal} {\bibinfo
   {journal} {Phys. Rev. B.}\ }\textbf {\bibinfo {volume} {104}},\ \bibinfo
  {pages} {075410} (\bibinfo {year} {2021})}\BibitemShut {NoStop}%
\bibitem [{\citenamefont {Wu}\ \emph {et~al.}(2019)\citenamefont {Wu},
  \citenamefont {Smart}, \citenamefont {Xu},\ and\ \citenamefont
  {Ping}}]{Wu2019}%
  \BibitemOpen
  \bibfield  {author} {\bibinfo {author} {\bibfnamefont {F.}~\bibnamefont
  {Wu}}, \bibinfo {author} {\bibfnamefont {T.~J.}\ \bibnamefont {Smart}},
  \bibinfo {author} {\bibfnamefont {J.}~\bibnamefont {Xu}},\ and\ \bibinfo
  {author} {\bibfnamefont {Y.}~\bibnamefont {Ping}},\ }\bibfield  {title}
  {\bibinfo {title} {Carrier recombination mechanism at defects in wide band
  gap two-dimensional materials from first principles},\ }\href
  {https://doi.org/10.1103/PhysRevB.100.081407} {\bibfield  {journal} {\bibinfo
   {journal} {Phys. Rev. B.}\ }\textbf {\bibinfo {volume} {100}},\ \bibinfo
  {pages} {081407} (\bibinfo {year} {2019})}\BibitemShut {NoStop}%
\bibitem [{\citenamefont {Sajid}\ and\ \citenamefont
  {Thygesen}(2020)}]{Sajid2020}%
  \BibitemOpen
  \bibfield  {author} {\bibinfo {author} {\bibfnamefont {A.}~\bibnamefont
  {Sajid}}\ and\ \bibinfo {author} {\bibfnamefont {K.~S.}\ \bibnamefont
  {Thygesen}},\ }\bibfield  {title} {\bibinfo {title}
  {V$_\text{N}$\text{C}$_\text{B}$ defect as source of single photon emission
  from hexagonal boron nitride},\ }\href
  {https://doi.org/10.1088/2053-1583/ab8f61} {\bibfield  {journal} {\bibinfo
  {journal} {2D Mater.}\ }\textbf {\bibinfo {volume} {7}},\ \bibinfo {pages}
  {031007} (\bibinfo {year} {2020})}\BibitemShut {NoStop}%
\bibitem [{\citenamefont {Jara}\ \emph {et~al.}(2021)\citenamefont {Jara},
  \citenamefont {Rauch}, \citenamefont {Botti}, \citenamefont {Marques},
  \citenamefont {Norambuena}, \citenamefont {Coto}, \citenamefont
  {Castellanos-Águila}, \citenamefont {Maze},\ and\ \citenamefont
  {Munoz}}]{Jara2020}%
  \BibitemOpen
  \bibfield  {author} {\bibinfo {author} {\bibfnamefont {C.}~\bibnamefont
  {Jara}}, \bibinfo {author} {\bibfnamefont {T.}~\bibnamefont {Rauch}},
  \bibinfo {author} {\bibfnamefont {S.}~\bibnamefont {Botti}}, \bibinfo
  {author} {\bibfnamefont {M.~A.~L.}\ \bibnamefont {Marques}}, \bibinfo
  {author} {\bibfnamefont {A.}~\bibnamefont {Norambuena}}, \bibinfo {author}
  {\bibfnamefont {R.}~\bibnamefont {Coto}}, \bibinfo {author} {\bibfnamefont
  {J.~E.}\ \bibnamefont {Castellanos-Águila}}, \bibinfo {author}
  {\bibfnamefont {J.~R.}\ \bibnamefont {Maze}},\ and\ \bibinfo {author}
  {\bibfnamefont {F.}~\bibnamefont {Munoz}},\ }\bibfield  {title} {\bibinfo
  {title} {First-principles identification of single photon emitters based on
  carbon clusters in hexagonal boron nitride},\ }\href
  {https://doi.org/10.1021/acs.jpca.0c07339} {\bibfield  {journal} {\bibinfo
  {journal} {J. Phys. Chem. A}\ }\textbf {\bibinfo {volume} {125}},\ \bibinfo
  {pages} {1325–1335} (\bibinfo {year} {2021})}\BibitemShut {NoStop}%
\bibitem [{\citenamefont {Li}\ \emph {et~al.}(2022)\citenamefont {Li},
  \citenamefont {Smart},\ and\ \citenamefont {Ping}}]{Li2022}%
  \BibitemOpen
  \bibfield  {author} {\bibinfo {author} {\bibfnamefont {K.}~\bibnamefont
  {Li}}, \bibinfo {author} {\bibfnamefont {T.~J.}\ \bibnamefont {Smart}},\ and\
  \bibinfo {author} {\bibfnamefont {Y.}~\bibnamefont {Ping}},\ }\bibfield
  {title} {\bibinfo {title} {Carbon trimer as a 2 ev single-photon emitter
  candidate in hexagonal boron nitride: A first-principles study},\ }\href
  {https://doi.org/10.1103/PhysRevMaterials.6.L042201} {\bibfield  {journal}
  {\bibinfo  {journal} {Phys. Rev. Mater.}\ }\textbf {\bibinfo {volume} {6}},\
  \bibinfo {pages} {L042201} (\bibinfo {year} {2022})}\BibitemShut {NoStop}%
\bibitem [{\citenamefont {Smart}\ \emph {et~al.}(2021)\citenamefont {Smart},
  \citenamefont {Li}, \citenamefont {Xu},\ and\ \citenamefont
  {Ping}}]{Smart2021}%
  \BibitemOpen
  \bibfield  {author} {\bibinfo {author} {\bibfnamefont {T.~J.}\ \bibnamefont
  {Smart}}, \bibinfo {author} {\bibfnamefont {K.}~\bibnamefont {Li}}, \bibinfo
  {author} {\bibfnamefont {J.}~\bibnamefont {Xu}},\ and\ \bibinfo {author}
  {\bibfnamefont {Y.}~\bibnamefont {Ping}},\ }\bibfield  {title} {\bibinfo
  {title} {Intersystem crossing and exciton–defect coupling of spin defects
  in hexagonal boron nitride},\ }\href
  {https://doi.org/10.1038/s41524-021-00525-5} {\bibfield  {journal} {\bibinfo
  {journal} {Npj Comput. Mater.}\ }\textbf {\bibinfo {volume} {7}},\ \bibinfo
  {pages} {59} (\bibinfo {year} {2021})}\BibitemShut {NoStop}%
\bibitem [{\citenamefont {Sajid}\ \emph
  {et~al.}(2018{\natexlab{a}})\citenamefont {Sajid}, \citenamefont {Reimers},\
  and\ \citenamefont {Ford}}]{Sajid2018}%
  \BibitemOpen
  \bibfield  {author} {\bibinfo {author} {\bibfnamefont {A.}~\bibnamefont
  {Sajid}}, \bibinfo {author} {\bibfnamefont {J.~R.}\ \bibnamefont {Reimers}},\
  and\ \bibinfo {author} {\bibfnamefont {M.~J.}\ \bibnamefont {Ford}},\
  }\bibfield  {title} {\bibinfo {title} {Defect states in hexagonal boron
  nitride: Assignments of observed properties and prediction of properties
  relevant to quantum computation},\ }\href
  {https://doi.org/10.1103/PhysRevB.97.064101} {\bibfield  {journal} {\bibinfo
  {journal} {Phys. Rev. B.}\ }\textbf {\bibinfo {volume} {97}},\ \bibinfo
  {pages} {1} (\bibinfo {year} {2018}{\natexlab{a}})}\BibitemShut {NoStop}%
\bibitem [{\citenamefont {Ivády}\ \emph {et~al.}(2020)\citenamefont {Ivády},
  \citenamefont {Barcza}, \citenamefont {Thiering}, \citenamefont {Li},
  \citenamefont {Hamdi}, \citenamefont {Chou}, \citenamefont {Örs Legeza},\
  and\ \citenamefont {Gali}}]{VB-1-Gali}%
  \BibitemOpen
  \bibfield  {author} {\bibinfo {author} {\bibfnamefont {V.}~\bibnamefont
  {Ivády}}, \bibinfo {author} {\bibfnamefont {G.}~\bibnamefont {Barcza}},
  \bibinfo {author} {\bibfnamefont {G.}~\bibnamefont {Thiering}}, \bibinfo
  {author} {\bibfnamefont {S.}~\bibnamefont {Li}}, \bibinfo {author}
  {\bibfnamefont {H.}~\bibnamefont {Hamdi}}, \bibinfo {author} {\bibfnamefont
  {J.~P.}\ \bibnamefont {Chou}}, \bibinfo {author} {\bibnamefont {Örs
  Legeza}},\ and\ \bibinfo {author} {\bibfnamefont {A.}~\bibnamefont {Gali}},\
  }\bibfield  {title} {\bibinfo {title} {Ab initio theory of the negatively
  charged boron vacancy qubit in hexagonal boron nitride},\ }\href
  {https://doi.org/10.1038/s41524-020-0305-x} {\bibfield  {journal} {\bibinfo
  {journal} {Npj Comput. Mater.}\ }\textbf {\bibinfo {volume} {6}},\ \bibinfo
  {pages} {41} (\bibinfo {year} {2020})}\BibitemShut {NoStop}%
\bibitem [{\citenamefont {Li}\ and\ \citenamefont
  {Gali}(2022)}]{Li2022-oxygen}%
  \BibitemOpen
  \bibfield  {author} {\bibinfo {author} {\bibfnamefont {S.}~\bibnamefont
  {Li}}\ and\ \bibinfo {author} {\bibfnamefont {A.}~\bibnamefont {Gali}},\
  }\bibfield  {title} {\bibinfo {title} {Identification of an oxygen defect in
  hexagonal boron nitride},\ }\href
  {https://doi.org/10.1021/acs.jpclett.2c02687} {\bibfield  {journal} {\bibinfo
   {journal} {J. Phys. Chem. Lett.}\ }\textbf {\bibinfo {volume} {13}},\
  \bibinfo {pages} {9544–9551} (\bibinfo {year} {2022})}\BibitemShut
  {NoStop}%
\bibitem [{\citenamefont {Sajid}\ \emph
  {et~al.}(2018{\natexlab{b}})\citenamefont {Sajid}, \citenamefont {Reimers},\
  and\ \citenamefont {Ford}}]{Sajid2018-EPR}%
  \BibitemOpen
  \bibfield  {author} {\bibinfo {author} {\bibfnamefont {A.}~\bibnamefont
  {Sajid}}, \bibinfo {author} {\bibfnamefont {J.~R.}\ \bibnamefont {Reimers}},\
  and\ \bibinfo {author} {\bibfnamefont {M.~J.}\ \bibnamefont {Ford}},\
  }\bibfield  {title} {\bibinfo {title} {Defect states in hexagonal boron
  nitride: Assignments of observed properties and prediction of properties
  relevant to quantum computation},\ }\href
  {https://doi.org/10.1103/PhysRevB.97.064101} {\bibfield  {journal} {\bibinfo
  {journal} {Phys. Rev. B.}\ }\textbf {\bibinfo {volume} {97}},\ \bibinfo
  {pages} {1} (\bibinfo {year} {2018}{\natexlab{b}})}\BibitemShut {NoStop}%
\bibitem [{\citenamefont {Kumar}\ \emph
  {et~al.}(2023{\natexlab{b}})\citenamefont {Kumar}, \citenamefont {Cholsuk},
  \citenamefont {Zand}, \citenamefont {Mishuk}, \citenamefont {Matthes},
  \citenamefont {Eilenberger}, \citenamefont {Suwanna},\ and\ \citenamefont
  {Vogl}}]{anand-yellow}%
  \BibitemOpen
  \bibfield  {author} {\bibinfo {author} {\bibfnamefont {A.}~\bibnamefont
  {Kumar}}, \bibinfo {author} {\bibfnamefont {C.}~\bibnamefont {Cholsuk}},
  \bibinfo {author} {\bibfnamefont {A.}~\bibnamefont {Zand}}, \bibinfo {author}
  {\bibfnamefont {M.~N.}\ \bibnamefont {Mishuk}}, \bibinfo {author}
  {\bibfnamefont {T.}~\bibnamefont {Matthes}}, \bibinfo {author} {\bibfnamefont
  {F.}~\bibnamefont {Eilenberger}}, \bibinfo {author} {\bibfnamefont
  {S.}~\bibnamefont {Suwanna}},\ and\ \bibinfo {author} {\bibfnamefont
  {T.}~\bibnamefont {Vogl}},\ }\bibfield  {title} {\bibinfo {title} {Localized
  creation of yellow single photon emitting carbon complexes in hexagonal boron
  nitride},\ }\bibfield  {journal} {\bibinfo  {journal} {APL Mater.}\ }\textbf
  {\bibinfo {volume} {11}},\ \href {https://doi.org/10.1063/5.0147560}
  {10.1063/5.0147560} (\bibinfo {year} {2023}{\natexlab{b}})\BibitemShut
  {NoStop}%
\bibitem [{\citenamefont {Kresse}\ and\ \citenamefont
  {Furthmüller}(1996)}]{vasp1}%
  \BibitemOpen
  \bibfield  {author} {\bibinfo {author} {\bibfnamefont {G.}~\bibnamefont
  {Kresse}}\ and\ \bibinfo {author} {\bibfnamefont {J.}~\bibnamefont
  {Furthmüller}},\ }\bibfield  {title} {\bibinfo {title} {Efficiency of
  ab-initio total energy calculations for metals and semiconductors using a
  plane-wave basis set},\ }\href
  {https://doi.org/https://doi.org/10.1016/0927-0256(96)00008-0} {\bibfield
  {journal} {\bibinfo  {journal} {Comput. Mater. Sci.}\ }\textbf {\bibinfo
  {volume} {6}},\ \bibinfo {pages} {15 } (\bibinfo {year} {1996})}\BibitemShut
  {NoStop}%
\bibitem [{\citenamefont {Kresse}\ and\ \citenamefont
  {Furthm\"uller}(1996)}]{vasp2}%
  \BibitemOpen
  \bibfield  {author} {\bibinfo {author} {\bibfnamefont {G.}~\bibnamefont
  {Kresse}}\ and\ \bibinfo {author} {\bibfnamefont {J.}~\bibnamefont
  {Furthm\"uller}},\ }\bibfield  {title} {\bibinfo {title} {Efficient iterative
  schemes for ab initio total-energy calculations using a plane-wave basis
  set},\ }\href {https://doi.org/https://doi.org/10.1103/PhysRevB.54.11169}
  {\bibfield  {journal} {\bibinfo  {journal} {Phys. Rev. B}\ }\textbf {\bibinfo
  {volume} {54}},\ \bibinfo {pages} {11169} (\bibinfo {year}
  {1996})}\BibitemShut {NoStop}%
\bibitem [{\citenamefont {Bl\"ochl}(1994)}]{paw}%
  \BibitemOpen
  \bibfield  {author} {\bibinfo {author} {\bibfnamefont {P.~E.}\ \bibnamefont
  {Bl\"ochl}},\ }\bibfield  {title} {\bibinfo {title} {Projector augmented-wave
  method},\ }\href {https://doi.org/https://doi.org/10.1103/PhysRevB.50.17953}
  {\bibfield  {journal} {\bibinfo  {journal} {Phys. Rev. B}\ }\textbf {\bibinfo
  {volume} {50}},\ \bibinfo {pages} {17953} (\bibinfo {year}
  {1994})}\BibitemShut {NoStop}%
\bibitem [{\citenamefont {Kresse}\ and\ \citenamefont {Joubert}(1999)}]{paw2}%
  \BibitemOpen
  \bibfield  {author} {\bibinfo {author} {\bibfnamefont {G.}~\bibnamefont
  {Kresse}}\ and\ \bibinfo {author} {\bibfnamefont {D.}~\bibnamefont
  {Joubert}},\ }\bibfield  {title} {\bibinfo {title} {From ultrasoft
  pseudopotentials to the projector augmented-wave method},\ }\href
  {https://doi.org/https://doi.org/10.1103/PhysRevB.59.1758} {\bibfield
  {journal} {\bibinfo  {journal} {Phys. Rev. B}\ }\textbf {\bibinfo {volume}
  {59}},\ \bibinfo {pages} {1758} (\bibinfo {year} {1999})}\BibitemShut
  {NoStop}%
\bibitem [{\citenamefont {Abdi}\ \emph {et~al.}(2018)\citenamefont {Abdi},
  \citenamefont {Chou}, \citenamefont {Gali},\ and\ \citenamefont
  {Plenio}}]{Abdi2018}%
  \BibitemOpen
  \bibfield  {author} {\bibinfo {author} {\bibfnamefont {M.}~\bibnamefont
  {Abdi}}, \bibinfo {author} {\bibfnamefont {J.~P.}\ \bibnamefont {Chou}},
  \bibinfo {author} {\bibfnamefont {A.}~\bibnamefont {Gali}},\ and\ \bibinfo
  {author} {\bibfnamefont {M.~B.}\ \bibnamefont {Plenio}},\ }\bibfield  {title}
  {\bibinfo {title} {Color centers in hexagonal boron nitride monolayers: A
  group theory and ab initio analysis},\ }\href
  {https://doi.org/10.1021/acsphotonics.7b01442} {\bibfield  {journal}
  {\bibinfo  {journal} {ACS Photonics}\ }\textbf {\bibinfo {volume} {5}},\
  \bibinfo {pages} {1967} (\bibinfo {year} {2018})}\BibitemShut {NoStop}%
\bibitem [{\citenamefont {Jones}\ and\ \citenamefont
  {Gunnarsson}(1989)}]{Jones1989}%
  \BibitemOpen
  \bibfield  {author} {\bibinfo {author} {\bibfnamefont {R.~O.}\ \bibnamefont
  {Jones}}\ and\ \bibinfo {author} {\bibfnamefont {O.}~\bibnamefont
  {Gunnarsson}},\ }\bibfield  {title} {\bibinfo {title} {The density functional
  formalism, its applications and prospects},\ }\href
  {https://doi.org/10.1103/RevModPhys.61.689} {\bibfield  {journal} {\bibinfo
  {journal} {Rev. Mod. Phys.}\ }\textbf {\bibinfo {volume} {61}},\ \bibinfo
  {pages} {689} (\bibinfo {year} {1989})}\BibitemShut {NoStop}%
\bibitem [{\citenamefont {Reimers}\ \emph {et~al.}(2020)\citenamefont
  {Reimers}, \citenamefont {Shen}, \citenamefont {Kianinia}, \citenamefont
  {Bradac}, \citenamefont {Aharonovich}, \citenamefont {Ford},\ and\
  \citenamefont {Piecuch}}]{Reimers2020-VB-1}%
  \BibitemOpen
  \bibfield  {author} {\bibinfo {author} {\bibfnamefont {J.~R.}\ \bibnamefont
  {Reimers}}, \bibinfo {author} {\bibfnamefont {J.}~\bibnamefont {Shen}},
  \bibinfo {author} {\bibfnamefont {M.}~\bibnamefont {Kianinia}}, \bibinfo
  {author} {\bibfnamefont {C.}~\bibnamefont {Bradac}}, \bibinfo {author}
  {\bibfnamefont {I.}~\bibnamefont {Aharonovich}}, \bibinfo {author}
  {\bibfnamefont {M.~J.}\ \bibnamefont {Ford}},\ and\ \bibinfo {author}
  {\bibfnamefont {P.}~\bibnamefont {Piecuch}},\ }\bibfield  {title} {\bibinfo
  {title} {Photoluminescence, photophysics, and photochemistry of the
  v$_\text{B}^{-1}$ defect in hexagonal boron nitride},\ }\bibfield  {journal}
  {\bibinfo  {journal} {Phys. Rev. B}\ }\textbf {\bibinfo {volume} {102}},\
  \href {https://doi.org/10.1103/PhysRevB.102.144105}
  {10.1103/PhysRevB.102.144105} (\bibinfo {year} {2020})\BibitemShut {NoStop}%
\bibitem [{\citenamefont {Reimers}\ \emph {et~al.}(2018)\citenamefont
  {Reimers}, \citenamefont {Sajid}, \citenamefont {Kobayashi},\ and\
  \citenamefont {Ford}}]{doi:10.1021/acs.jctc.7b01072}%
  \BibitemOpen
  \bibfield  {author} {\bibinfo {author} {\bibfnamefont {J.~R.}\ \bibnamefont
  {Reimers}}, \bibinfo {author} {\bibfnamefont {A.}~\bibnamefont {Sajid}},
  \bibinfo {author} {\bibfnamefont {R.}~\bibnamefont {Kobayashi}},\ and\
  \bibinfo {author} {\bibfnamefont {M.~J.}\ \bibnamefont {Ford}},\ }\bibfield
  {title} {\bibinfo {title} {Understanding and calibrating
  density-functional-theory calculations describing the energy and spectroscopy
  of defect sites in hexagonal boron nitride},\ }\href
  {https://doi.org/10.1021/acs.jctc.7b01072} {\bibfield  {journal} {\bibinfo
  {journal} {J. Chem. Theory Comput.}\ }\textbf {\bibinfo {volume} {14}},\
  \bibinfo {pages} {1602} (\bibinfo {year} {2018})}\BibitemShut {NoStop}%
\bibitem [{\citenamefont {Liu}(1 01)}]{pyWave}%
  \BibitemOpen
  \bibfield  {author} {\bibinfo {author} {\bibfnamefont {L.}~\bibnamefont
  {Liu}},\ }\href {https://github.com/liming-liu/pyvaspwfc} {\bibinfo {title}
  {Pyvaspwfc. https://github.com/liming-liu/pyvaspwfc}} (\bibinfo {year}
  {accessed 2022-11-01})\BibitemShut {NoStop}%
\bibitem [{\citenamefont {Davidsson}(2020)}]{Davidsson2020}%
  \BibitemOpen
  \bibfield  {author} {\bibinfo {author} {\bibfnamefont {J.}~\bibnamefont
  {Davidsson}},\ }\bibfield  {title} {\bibinfo {title} {Theoretical
  polarization of zero phonon lines in point defects},\ }\href
  {https://doi.org/https://doi.org/10.1088/1361-648X/ab94f4} {\bibfield
  {journal} {\bibinfo  {journal} {J. Phys.: Condens. Matter}\ }\textbf
  {\bibinfo {volume} {32}},\ \bibinfo {pages} {385502} (\bibinfo {year}
  {2020})}\BibitemShut {NoStop}%
\bibitem [{\citenamefont {Turiansky}\ \emph {et~al.}(2021)\citenamefont
  {Turiansky}, \citenamefont {Alkauskas}, \citenamefont {Engel}, \citenamefont
  {Kresse}, \citenamefont {Wickramaratne}, \citenamefont {Shen}, \citenamefont
  {Dreyer},\ and\ \citenamefont {de~Walle}}]{Turiansky2021}%
  \BibitemOpen
  \bibfield  {author} {\bibinfo {author} {\bibfnamefont {M.~E.}\ \bibnamefont
  {Turiansky}}, \bibinfo {author} {\bibfnamefont {A.}~\bibnamefont
  {Alkauskas}}, \bibinfo {author} {\bibfnamefont {M.}~\bibnamefont {Engel}},
  \bibinfo {author} {\bibfnamefont {G.}~\bibnamefont {Kresse}}, \bibinfo
  {author} {\bibfnamefont {D.}~\bibnamefont {Wickramaratne}}, \bibinfo {author}
  {\bibfnamefont {J.~X.}\ \bibnamefont {Shen}}, \bibinfo {author}
  {\bibfnamefont {C.~E.}\ \bibnamefont {Dreyer}},\ and\ \bibinfo {author}
  {\bibfnamefont {C.~G.~V.}\ \bibnamefont {de~Walle}},\ }\bibfield  {title}
  {\bibinfo {title} {Nonrad: Computing nonradiative capture coefficients from
  first principles},\ }\href {https://doi.org/10.1016/j.cpc.2021.108056}
  {\bibfield  {journal} {\bibinfo  {journal} {Comput. Phys. Commu.}\ }\textbf
  {\bibinfo {volume} {267}},\ \bibinfo {pages} {108056} (\bibinfo {year}
  {2021})}\BibitemShut {NoStop}%
\bibitem [{\citenamefont {Alkauskas}\ \emph
  {et~al.}(2014{\natexlab{a}})\citenamefont {Alkauskas}, \citenamefont {Yan},\
  and\ \citenamefont {Van~de Walle}}]{alkauskas_first-principles_2014}%
  \BibitemOpen
  \bibfield  {author} {\bibinfo {author} {\bibfnamefont {A.}~\bibnamefont
  {Alkauskas}}, \bibinfo {author} {\bibfnamefont {Q.}~\bibnamefont {Yan}},\
  and\ \bibinfo {author} {\bibfnamefont {C.~G.}\ \bibnamefont {Van~de Walle}},\
  }\bibfield  {title} {\bibinfo {title} {First-principles theory of
  nonradiative carrier capture via multiphonon emission},\ }\href
  {https://doi.org/10.1103/PhysRevB.90.075202} {\bibfield  {journal} {\bibinfo
  {journal} {Phys. Rev. B}\ }\textbf {\bibinfo {volume} {90}},\ \bibinfo
  {pages} {075202} (\bibinfo {year} {2014}{\natexlab{a}})}\BibitemShut
  {NoStop}%
\bibitem [{\citenamefont {Tawfik}\ and\ \citenamefont
  {Russo}(2022)}]{Tawfik2022}%
  \BibitemOpen
  \bibfield  {author} {\bibinfo {author} {\bibfnamefont {S.~A.}\ \bibnamefont
  {Tawfik}}\ and\ \bibinfo {author} {\bibfnamefont {S.~P.}\ \bibnamefont
  {Russo}},\ }\bibfield  {title} {\bibinfo {title} {Pyphotonics: A python
  package for the evaluation of luminescence properties of defects},\ }\href
  {https://doi.org/10.1016/j.cpc.2021.108222} {\bibfield  {journal} {\bibinfo
  {journal} {Comput. Phys. Commun.}\ }\textbf {\bibinfo {volume} {273}},\
  \bibinfo {pages} {108222} (\bibinfo {year} {2022})}\BibitemShut {NoStop}%
\bibitem [{\citenamefont {Alkauskas}\ \emph
  {et~al.}(2014{\natexlab{b}})\citenamefont {Alkauskas}, \citenamefont
  {Buckley}, \citenamefont {Awschalom},\ and\ \citenamefont
  {Walle}}]{Alkauskas2014}%
  \BibitemOpen
  \bibfield  {author} {\bibinfo {author} {\bibfnamefont {A.}~\bibnamefont
  {Alkauskas}}, \bibinfo {author} {\bibfnamefont {B.~B.}\ \bibnamefont
  {Buckley}}, \bibinfo {author} {\bibfnamefont {D.~D.}\ \bibnamefont
  {Awschalom}},\ and\ \bibinfo {author} {\bibfnamefont {C.~G. V.~D.}\
  \bibnamefont {Walle}},\ }\bibfield  {title} {\bibinfo {title}
  {First-principles theory of the luminescence lineshape for the triplet
  transition in diamond nv centres},\ }\href
  {https://doi.org/10.1088/1367-2630/16/7/073026} {\bibfield  {journal}
  {\bibinfo  {journal} {New J. Phys.}\ }\textbf {\bibinfo {volume} {16}},\
  \bibinfo {pages} {073026} (\bibinfo {year} {2014}{\natexlab{b}})}\BibitemShut
  {NoStop}%
\bibitem [{\citenamefont {Mendelson}\ \emph {et~al.}(2021)\citenamefont
  {Mendelson}, \citenamefont {Chugh}, \citenamefont {Reimers}, \citenamefont
  {Cheng}, \citenamefont {Gottscholl}, \citenamefont {Long}, \citenamefont
  {Mellor}, \citenamefont {Zettl}, \citenamefont {Dyakonov}, \citenamefont
  {Beton}, \citenamefont {Novikov}, \citenamefont {Jagadish}, \citenamefont
  {Tan}, \citenamefont {Ford}, \citenamefont {Toth}, \citenamefont {Bradac},\
  and\ \citenamefont {Aharonovich}}]{Mendelson2021}%
  \BibitemOpen
  \bibfield  {author} {\bibinfo {author} {\bibfnamefont {N.}~\bibnamefont
  {Mendelson}}, \bibinfo {author} {\bibfnamefont {D.}~\bibnamefont {Chugh}},
  \bibinfo {author} {\bibfnamefont {J.~R.}\ \bibnamefont {Reimers}}, \bibinfo
  {author} {\bibfnamefont {T.~S.}\ \bibnamefont {Cheng}}, \bibinfo {author}
  {\bibfnamefont {A.}~\bibnamefont {Gottscholl}}, \bibinfo {author}
  {\bibfnamefont {H.}~\bibnamefont {Long}}, \bibinfo {author} {\bibfnamefont
  {C.~J.}\ \bibnamefont {Mellor}}, \bibinfo {author} {\bibfnamefont
  {A.}~\bibnamefont {Zettl}}, \bibinfo {author} {\bibfnamefont
  {V.}~\bibnamefont {Dyakonov}}, \bibinfo {author} {\bibfnamefont {P.~H.}\
  \bibnamefont {Beton}}, \bibinfo {author} {\bibfnamefont {S.~V.}\ \bibnamefont
  {Novikov}}, \bibinfo {author} {\bibfnamefont {C.}~\bibnamefont {Jagadish}},
  \bibinfo {author} {\bibfnamefont {H.~H.}\ \bibnamefont {Tan}}, \bibinfo
  {author} {\bibfnamefont {M.~J.}\ \bibnamefont {Ford}}, \bibinfo {author}
  {\bibfnamefont {M.}~\bibnamefont {Toth}}, \bibinfo {author} {\bibfnamefont
  {C.}~\bibnamefont {Bradac}},\ and\ \bibinfo {author} {\bibfnamefont
  {I.}~\bibnamefont {Aharonovich}},\ }\bibfield  {title} {\bibinfo {title}
  {Identifying carbon as the source of visible single-photon emission from
  hexagonal boron nitride},\ }\href
  {https://doi.org/10.1038/s41563-020-00850-y} {\bibfield  {journal} {\bibinfo
  {journal} {Nat. Mater.}\ }\textbf {\bibinfo {volume} {20}},\ \bibinfo {pages}
  {321} (\bibinfo {year} {2021})}\BibitemShut {NoStop}%
\bibitem [{\citenamefont {Wang}\ \emph {et~al.}(2018)\citenamefont {Wang},
  \citenamefont {Zhang}, \citenamefont {Zhao}, \citenamefont {Luo},
  \citenamefont {Wong}, \citenamefont {Wang}, \citenamefont {Wan},
  \citenamefont {Venkatesan}, \citenamefont {Pennycook}, \citenamefont {Loh},
  \citenamefont {Eda},\ and\ \citenamefont {Wee}}]{Wang2018}%
  \BibitemOpen
  \bibfield  {author} {\bibinfo {author} {\bibfnamefont {Q.}~\bibnamefont
  {Wang}}, \bibinfo {author} {\bibfnamefont {Q.}~\bibnamefont {Zhang}},
  \bibinfo {author} {\bibfnamefont {X.}~\bibnamefont {Zhao}}, \bibinfo {author}
  {\bibfnamefont {X.}~\bibnamefont {Luo}}, \bibinfo {author} {\bibfnamefont
  {C.~P.~Y.}\ \bibnamefont {Wong}}, \bibinfo {author} {\bibfnamefont
  {J.}~\bibnamefont {Wang}}, \bibinfo {author} {\bibfnamefont {D.}~\bibnamefont
  {Wan}}, \bibinfo {author} {\bibfnamefont {T.}~\bibnamefont {Venkatesan}},
  \bibinfo {author} {\bibfnamefont {S.~J.}\ \bibnamefont {Pennycook}}, \bibinfo
  {author} {\bibfnamefont {K.~P.}\ \bibnamefont {Loh}}, \bibinfo {author}
  {\bibfnamefont {G.}~\bibnamefont {Eda}},\ and\ \bibinfo {author}
  {\bibfnamefont {A.~T.~S.}\ \bibnamefont {Wee}},\ }\bibfield  {title}
  {\bibinfo {title} {Photoluminescence upconversion by defects in hexagonal
  boron nitride},\ }\href {https://doi.org/10.1021/acs.nanolett.8b02804}
  {\bibfield  {journal} {\bibinfo  {journal} {Nano Lett.}\ }\textbf {\bibinfo
  {volume} {18}},\ \bibinfo {pages} {6898} (\bibinfo {year}
  {2018})}\BibitemShut {NoStop}%
\bibitem [{\citenamefont {Exarhos}\ \emph {et~al.}(2017)\citenamefont
  {Exarhos}, \citenamefont {Hopper}, \citenamefont {Grote}, \citenamefont
  {Alkauskas},\ and\ \citenamefont {Bassett}}]{Exarhos2017}%
  \BibitemOpen
  \bibfield  {author} {\bibinfo {author} {\bibfnamefont {A.~L.}\ \bibnamefont
  {Exarhos}}, \bibinfo {author} {\bibfnamefont {D.~A.}\ \bibnamefont {Hopper}},
  \bibinfo {author} {\bibfnamefont {R.~R.}\ \bibnamefont {Grote}}, \bibinfo
  {author} {\bibfnamefont {A.}~\bibnamefont {Alkauskas}},\ and\ \bibinfo
  {author} {\bibfnamefont {L.~C.}\ \bibnamefont {Bassett}},\ }\bibfield
  {title} {\bibinfo {title} {Optical signatures of quantum emitters in
  suspended hexagonal boron nitride},\ }\href
  {https://doi.org/10.1021/acsnano.7b00665} {\bibfield  {journal} {\bibinfo
  {journal} {ACS Nano}\ }\textbf {\bibinfo {volume} {11}},\ \bibinfo {pages}
  {3328} (\bibinfo {year} {2017})}\BibitemShut {NoStop}%
\bibitem [{\citenamefont {Kozawa}\ \emph {et~al.}(2023)\citenamefont {Kozawa},
  \citenamefont {Li}, \citenamefont {Ichihara}, \citenamefont {Rajan},
  \citenamefont {Gong}, \citenamefont {He}, \citenamefont {Koman},
  \citenamefont {Zeng}, \citenamefont {Kuehne}, \citenamefont {Silmore},
  \citenamefont {Parviz}, \citenamefont {Liu}, \citenamefont {Liu},
  \citenamefont {Faucher}, \citenamefont {Yuan}, \citenamefont {Warner},
  \citenamefont {Blankschtein},\ and\ \citenamefont {Strano}}]{Kozawa2023}%
  \BibitemOpen
  \bibfield  {author} {\bibinfo {author} {\bibfnamefont {D.}~\bibnamefont
  {Kozawa}}, \bibinfo {author} {\bibfnamefont {S.~X.}\ \bibnamefont {Li}},
  \bibinfo {author} {\bibfnamefont {T.}~\bibnamefont {Ichihara}}, \bibinfo
  {author} {\bibfnamefont {A.~G.}\ \bibnamefont {Rajan}}, \bibinfo {author}
  {\bibfnamefont {X.}~\bibnamefont {Gong}}, \bibinfo {author} {\bibfnamefont
  {G.}~\bibnamefont {He}}, \bibinfo {author} {\bibfnamefont {V.~B.}\
  \bibnamefont {Koman}}, \bibinfo {author} {\bibfnamefont {Y.}~\bibnamefont
  {Zeng}}, \bibinfo {author} {\bibfnamefont {M.}~\bibnamefont {Kuehne}},
  \bibinfo {author} {\bibfnamefont {K.~S.}\ \bibnamefont {Silmore}}, \bibinfo
  {author} {\bibfnamefont {D.}~\bibnamefont {Parviz}}, \bibinfo {author}
  {\bibfnamefont {P.}~\bibnamefont {Liu}}, \bibinfo {author} {\bibfnamefont
  {A.~T.}\ \bibnamefont {Liu}}, \bibinfo {author} {\bibfnamefont
  {S.}~\bibnamefont {Faucher}}, \bibinfo {author} {\bibfnamefont
  {Z.}~\bibnamefont {Yuan}}, \bibinfo {author} {\bibfnamefont {J.}~\bibnamefont
  {Warner}}, \bibinfo {author} {\bibfnamefont {D.}~\bibnamefont
  {Blankschtein}},\ and\ \bibinfo {author} {\bibfnamefont {M.~S.}\ \bibnamefont
  {Strano}},\ }\bibfield  {title} {\bibinfo {title} {Discretized hexagonal
  boron nitride quantum emitters and their chemical interconversion},\ }\href
  {https://doi.org/10.1088/1361-6528/aca984} {\bibfield  {journal} {\bibinfo
  {journal} {Nanotechnology}\ }\textbf {\bibinfo {volume} {34}},\ \bibinfo
  {pages} {115702} (\bibinfo {year} {2023})}\BibitemShut {NoStop}%
\bibitem [{\citenamefont {Jungwirth}\ and\ \citenamefont
  {Fuchs}(2017)}]{Jungwirth2017}%
  \BibitemOpen
  \bibfield  {author} {\bibinfo {author} {\bibfnamefont {N.~R.}\ \bibnamefont
  {Jungwirth}}\ and\ \bibinfo {author} {\bibfnamefont {G.~D.}\ \bibnamefont
  {Fuchs}},\ }\bibfield  {title} {\bibinfo {title} {Optical absorption and
  emission mechanisms of single defects in hexagonal boron nitride},\ }\href
  {https://doi.org/10.1103/PhysRevLett.119.057401} {\bibfield  {journal}
  {\bibinfo  {journal} {Phys. Rev. Lett.}\ }\textbf {\bibinfo {volume} {119}},\
  \bibinfo {pages} {057401} (\bibinfo {year} {2017})}\BibitemShut {NoStop}%
\end{thebibliography}%

\clearpage

\end{document}